\documentclass[a4paper,fleqn,usenatbib]{mnras}

\usepackage[T1]{fontenc}
\usepackage{ae,aecompl}

\usepackage{graphicx}	% Including figure files
\usepackage{amsmath}	% Advanced maths commands
\usepackage{amssymb}	% Extra maths symbols
\usepackage{breakurl}

\newcommand{\ditto}{$- \prime \prime -$}
\newcommand{\dg}{^{\circ}}

\title[RoboPol: EVPA rotations in blazars]{RoboPol: first season rotations of optical polarization plane in blazars}

\author[D. Blinov et al.]
{D. Blinov$^{1,7}$\thanks{E-mail: blinov@physics.uoc.gr}, V. Pavlidou$^{1,2}$, I. Papadakis$^{1,2}$,
S. Kiehlmann$^3$, G. Panopoulou$^1$,
\newauthor
I. Liodakis$^1$, O.\,G. King$^{4}$, E. Angelakis$^3$, M. Balokovi\'{c}$^4$, H. Das$^6$, R. Feiler$^5$,
\newauthor
L. Fuhrmann$^3$, T. Hovatta$^{8}$, P. Khodade$^6$, A. Kus$^5$, N. Kylafis$^{2,1}$, A. Mahabal$^{4}$,
\newauthor
I. Myserlis$^{3}$, D. Modi$^{6}$, B. Pazderska$^5$, E. Pazderski$^5$,  I. Papamastorakis$^{1,2}$,
\newauthor
T.\,J. Pearson$^{4}$, C. Rajarshi$^6$, A. Ramaprakash$^6$, P. Reig$^{2,1}$, A.\,C.\,S. Readhead$^{4}$,
\newauthor
K. Tassis$^{1,2}$, J.\,A. Zensus$^3$ \\
$^{1}$Department of Physics and Institute for Plasma Physics, University of Crete, GR-71003, Heraklion, Greece\\
$^{2}$Foundation for Research and Technology - Hellas, IESL, Voutes, 7110 Heraklion, Greece\\
$^{3}$Max-Planck-Institut f\"{u}r Radioastronomie, Auf dem H\"{u}gel
69, 53121 Bonn, Germany\\
$^{4}$Cahill Center for Astronomy and Astrophysics, California Institute of Technology, 1200 E California Blvd,
MC 249-17,\\Pasadena CA, 
91125, USA\\
$^5$Toru\'{n} Centre for Astronomy, Nicolaus Copernicus University, Faculty of Physics, Astronomy and Informatics,\\
Grudziadzka 5, 87-100 Toru\'{n}, Poland\\
$^6$Inter-University Centre for Astronomy and Astrophysics, Post Bag
4, Ganeshkhind, Pune - 411 007, India\\
$^7$Astronomical Institute, St. Petersburg State University,Universitetsky pr. 28, Petrodvoretz, 198504 St. Petersburg,
Russia \\
$^8$Aalto University Mets\"ahovi Radio Observatory, Mets\"ahovintie 114, FL-02540 Kylm\"al\"a, Finland
}

\date{Accepted XXX. Received YYY; in original form ZZZ}

\pubyear{2015}

\begin{document}
\label{firstpage}
\pagerange{\pageref{firstpage}--\pageref{lastpage}}
\maketitle

\begin{abstract}
We present first results on polarization swings in optical emission of blazars obtained by RoboPol, a monitoring programme
of an unbiased sample of gamma-ray bright blazars specially designed for effective detection of such events. A possible
connection of polarization swing events with periods of high activity in gamma rays is investigated using the data set
obtained during the first season of operation. It was found that the brightest gamma-ray flares tend to be located closer
in time to rotation events, which may be an indication of two separate mechanisms responsible for the rotations. Blazars
with detected rotations during non-rotating periods have significantly larger amplitude and faster variations of
polarization angle than blazars without rotations. Our simulations show that the full set of observed rotations is
not a likely outcome (probability $\le 1.5 \times 10^{-2}$) of a random walk of the polarization vector simulated by a
multicell model. Furthermore, it is highly unlikely ($\sim 5 \times 10^{-5}$) that none of our rotations is physically
connected with an increase in gamma-ray activity. 
\end{abstract}


%%%%%%%%%%%%%%%%%%%%%%%%%%%%%%%%%%%%%%%%%%%%%%%%%%

%%%%%%%%%%%%%%%%% BODY OF PAPER %%%%%%%%%%%%%%%%%%
\section{Introduction} \label{sec:introduction}

Blazars are active galactic nuclei whose jets are oriented close to our line of sight, so that we observe high relativistic beaming
of their non-thermal emission and large amplitude variability at all wavelengths. The low-frequency emission is dominated by
synchrotron radiation, and hence is highly polarized. The exact polarization fraction and direction depend on the structure
of the magnetic field in the emitting region, and on the number of emitting regions along the line of sight. The polarization
direction (in the simple case of a single dominant emission region) traces (and is perpendicular to) the direction of the
projected magnetic field on the plane of the sky. Already from early optical observations, it has been known that
polarization parameters of blazars are variable on daily time-scales \citep{Kinman1966}. In general, both flux density and
polarization exhibit an erratic variability \citep{Angel1980,Uemura2010,Ikejiri2011}, which could be interpreted as a
random walk \citep{Moore1982}.
\begin{table*}
 \centering
  \caption{\label{tab:sample_select} Selection criteria for the gamma-ray--loud and the control sample.}
  \begin{tabular}{|p{.3\textwidth}|p{.3\textwidth}|p{.3\textwidth}|} 
     \hline
  Property             & Gamma-ray--loud sample     & Control sample         \\
  \hline
2FGL                                   & included                               & not included     \\\\[-1.5ex]
2FGL $\mathrm{F(E > 100 \,MeV)}$       & $> 10^{-8}{\rm\,cm^{-2}\,s^{-1}}$      & $-$              \\\\[-1.5ex]
2FGL source class                      & agu, bzb, or bzq                       & $-$              \\\\[-1.5ex]
Galactic latitude $|b|$                & $> 10\dg$                           & $-$              \\\\[-1.5ex]
Elevation (Elv) constraints$^1$         &$\mathrm{Elv_{\rm max}}\geq 40\dg$ for at least 90 consecutive nights in the window June -- November& $\mathrm{Elv_{\rm max}}\geq 40\dg$ for at least 90 consecutive nights in the window April -- November\\\\[-1.5ex] 
$R$ magnitude$^{2}$                    & $\leq 17.5$                            & $\leq 17.5$     \\\\[-1.5ex]
CGRaBS/15\,GHz OVRO monitoring         &  no constraints                        & included        \\\\[-1.5ex]
OVRO 15\,GHz mean flux density         &  no constraints                        &$\geq 0.060$\,Jy \\\\[-1.5ex]
OVRO 15\,GHz intrinsic modulation index, $m$ & no constraints                   &$\geq 0.05$      \\\\
\hline
\end{tabular}
\begin{flushleft}                                                                                
$^1$Refers to elevation during Skinakas dark hours\\                                         
$^2$Average value between archival value and measured during preliminary RoboPol observations (when applicable)                                             
\end{flushleft}                                                                                  
\end{table*}
However, in some cases the electric vector position angle (EVPA) of the polarized emission
displays long, smooth and monotonic rotations which have been observed in the optical since the 1980s \citep{Kikuchi1988}.
A number of mechanisms have been proposed for the interpretation of such events, including: stochastic variations of turbulent
magnetic fields, a shock travelling through a non-axisymmetric magnetic field \citep{Konigl1985}, polarized flares
in the accretion disc \citep{Sillanpaa1993}, two-component models consisting of two independent sources of polarized
emission \citep{Bjornsson1982}, and jet bending \citep{Abdo2010}.

Blazars represent the most common class of known gamma-ray sources \citep{Nolan2012,Fermi2015}. Despite the recent
progress in the field, many questions concerning the high-energy emission produced by blazars are still under debate.
For instance, it is unclear where the gamma-ray emitting site is located: within the broad-line region
\citep[e.g.][]{Blandford1995,Poutanen2010} or well downstream in the jet \citep[e.g.][]{Marscher2008,Agudo2011}.

Recent work showed that at least some large EVPA swings can be associated with gamma-ray flares
\citep[e.g.][]{Abdo2010,Larionov2013} and therefore can possibly provide some insight on the physics of high-energy activity.
Although such events have triggered an increasing interest in polarimetric monitoring of gamma-ray blazars,
efforts in this direction have been based on selected cases comprising statistically biased samples. As a result, a
significant amount of invaluable polarimetric data sets for a large number of sources has been gathered. However, this set
cannot be used for statistically rigorous population studies and, in particular, the investigation of a possible
correlation between gamma-ray flares and optical EVPA rotations. The RoboPol programme \citep{King2014,Pavlidou2014} has been
designed to provide a data set of rotation events in an unbiased sample of blazars, appropriate for such studies.

In this paper, we analyse EVPA rotations detected by RoboPol during the first observing season between 2013 July and November.
After a brief description of observing and reduction techniques in Sec.~\ref{sec:observations}, we estimate
the frequency of EVPA rotations in blazars and list their properties in Sec.~\ref{sec:rotations}. A Monte Carlo
simulation is performed in Sec.~\ref{sec:random_walks} in order to determine whether the EVPA rotations can be produced by
random walk processes. In Sec.~\ref{sec:gamma} we study the possible connection between an increased activity in the
gamma-ray band and EVPA swings. Our findings are summarized in Sec.~\ref{sec:conclusion}.

\section{Observations and data reduction} \label{sec:observations}
\subsection{Our sample}

A unique feature of the RoboPol programme is that it is monitoring a sample which has been selected on the basis of strict,
bias-free and objective criteria \citep[for detailed discussion on the sample construction, see][]{Pavlidou2014}. The
sample consists of three distinct groups.
\begin{enumerate}
 \item The main (``gamma-ray--loud'') sample is an unbiased subset of a statistically
 complete flux-limited sample of blazars from the second \textit{Fermi}-LAT source catalogue \citep{Nolan2012}. Specifically,
 we selected all the sources in the 2FGL catalogue classified as BL Lacertae objects (bzb), Flat Spectrum Radio Quasars (bzq),
 or active galaxy of uncertain type (agu). Applying the selection criteria listed in Table~\ref{tab:sample_select}, we
 constructed a gamma-ray flux-limited ``parent sample''. Application of the visibility constraints and field-quality cuts
 resulted in an unbiased subsample of 83 sources, among which we randomly selected 62 sources.
 \item A ``control'' sample of 15 ``gamma-ray--quiet'' sources. It constitutes an unbiased subset of a statistically
 complete sample of blazars. It has been drawn from the CGRaBS catalogue \citep{Healey2008} applying the selection criteria
 listed in Table~\ref{tab:sample_select}.
 \item 24 additional sources chosen on the basis of their variability characteristics or their presence either in the F-GAMMA
 programme sample or in TeV catalogues.
\end{enumerate}
Although here we present the polarization swings detected in all monitored sources during the first RoboPol observing
season, the statistical analysis in this paper is based only on sources from the group (i).

\begin{figure}
 \centering
 \includegraphics[width=0.47\textwidth]{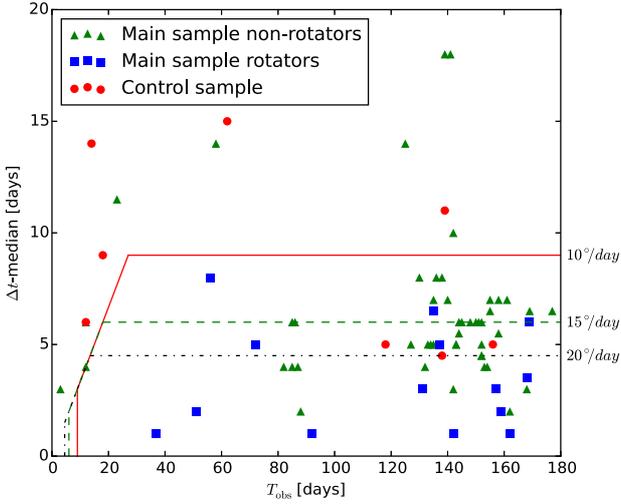}
\caption{Season length and median cadence for the first season data. Broken lines limit areas where rotations at rates of
10, 15 and 20 deg d$^{-1}$ can be detected (see Sec.~\ref{subsec:rot_freq} for details). Only objects with $\Delta t$-median
$\le 20$ were left for the more detailed view.}
 \label{fig:cadence_length}
\end{figure}

\subsection{Optical observations}
All photometric and polarimetric measurements were done at the 1.3-m telescope of Skinakas
observatory\footnote{\url{http://skinakas.physics.uoc.gr}} using RoboPol, a polarimeter specifically built for the project
\citep{King2014}. The RoboPol instrument contains a fixed set of two Wollaston prisms and half-wave plates,
which splits each incident ray into four rays with polarization plane rotated $45\dg$ with respect to each other.
Measuring relative intensities in pairs of the rays for each object in the 13 arcmin $\times$ 13 arcmin field, we obtain
the fractional Stokes parameters $q = (I_1-I_2)/(I_1+I_2) = Q/I$ and $u = (I_3-I_4)/(I_3+I_4) = U/I$. Stokes parameter $I$ is
calculated as a sum of intensities of all four spots. Since the polarization parameters are measured simultaneously, we
avoid unmeasurable errors caused by the sky changes between measurements and imperfect alignment of rotating optical elements.
\begin{table*}
\centering
\caption{Observational data for EVPA rotations detected by RoboPol in 2013. Columns (1),(2) - blazar identifiers;
(3) - redshift; (4) - observational season length; (5) - average rotation rate; (6) - total amplitude of EVPA change;
(7) - number of observations during rotation; (8) - time duration of the rotation;  (9) - TeV emission flag according to 
TeVCat$^{14}$ (``Y'' means that the blazar has been detected in gamma rays with $E>1$ TeV, ``N'' - otherwise);
(10) - blazar subclass (LBL, IBL, HBL denote low, intermediate and high synchrotron peaked BL Lacertae objects, FSRQ --
flat spectrum radio quasar).}
\label{tab:rbpl_rotations}
  \begin{tabular}{lccccccccc}
  \hline
 Blazar ID &   Survey      &$z$ & $T_{\rm obs}$ & $\langle \frac{\Delta \theta}{\Delta t} \rangle$ & $\Delta \theta_{\rm max}$ &  $N_{\rm points}$ & $T_{\rm rot}$ & TeV & Class  \\
           &   name        &    &   (d)         &                    (deg/d)                       &        (deg)              &                   &  (d)          &     &        \\
 \hline
RBPLJ0136+4751 & OC 457         & 0.859${}^1$   &  59  &  -6.6     & -225   & 6  & 34  &  N   & FSRQ${}^{13}$\\
RBPLJ0259+0747 & PKS0256+075    & 0.893${}^2$   &  72  &  -4.8     & -180   & 6  & 38  &  N   & FSRQ${}^{13}$\\
RBPLJ0721+7120* & S5 0716$+$71  & 0.31${}^{3}$  &  88  &  -14.8    & -208   & 11 & 14  &  Y   & LBL${}^{10}$ \\
RBPLJ0854+2006* & OJ 287        & 0.306${}^{4}$ &  51  &  -6.7     & -154   & 10 & 23  &  N   & LBL${}^{10}$  \\
RBPLJ1048+7143 & S5 1044$+$71   & 1.15${}^{5}$  &  142 &  -9.0     & -188   & 22 & 21  &  N   & $-$  \\
RBPLJ1555+1111 & PG 1553$+$113  &    $-$        &  129 &   5.6     &  128   & 8  & 23  &  Y   & HBL${}^{10}$ \\
RBPLJ1558+5625 & TXS1557$+$565  & 0.3${}^{6}$   &  137 &   7.2     &  222   & 9  & 31  &  N   & IBL?${}^{11}$  \\
RBPLJ1806+6949 & 3C 371         & 0.05${}^{7}$  &  143 &  -16.5    & -347   & 7  & 21  &  N   & LBL${}^{11}$  \\
RBPLJ1806+6949 & \ditto         & \ditto        &\ditto&   13.3    &  238   & 5  & 18  &  N   & \ditto     \\
RBPLJ1927+6117 & S4 1926+61     &    $-$        &  135 &   -4.4    & -105   & 6  & 24  &  N   & LBL${}^{13}$ \\
RBPLJ2202+4216 & BL Lac         & 0.069${}^{8}$ &  137 &   -51.0   & -253   & 5  & 5   &  Y   & LBL${}^{10}$  \\
RBPLJ2232+1143 & CTA 102        & 1.037${}^{1}$ &  140 &  -15.6    & -312   & 8  & 20  &  N   & FSRQ${}^{13}$ \\
RBPLJ2232+1143 & \ditto         & \ditto        &\ditto&  -11.8    & -140   & 6  & 12  &  N   & \ditto     \\
RBPLJ2243+2021 & RGB J2243+203  &    $-$        &  169 &  -5.9     & -183   & 5  & 31  &  N   & LBL${}^{12}$  \\
RBPLJ2253+1608 & 3C 454.3       & 0.859${}^{1}$ &  159 &  -18.3    & -129   & 4  & 7   &  N   & FSRQ${}^{13}$ \\
RBPLJ2311+3425 & B2 2308$+$34   & 1.817${}^{9}$ &  36  &  3.3      &  74    & 20 & 23  &  N   & FSRQ${}^{13}$ \\
\hline
\multicolumn{10}{l}{* Source belongs to sample (iii)}\\
\multicolumn{10}{l}{${}^1$\citep{Hewitt1987};${}^2$\citep{Murphy1993};${}^{3}$\citep{Nilsson2008};${}^{4}$\citep{Nilsson2010};}\\
\multicolumn{10}{l}{${}^{5}$\citep{Polatidis1995};${}^{6}$\citep{Falco1998};${}^{7}$\citep{deGrijp1992};${}^{8}$\citep{Vermeulen1995};}\\
\multicolumn{10}{l}{${}^{9}$\citep{Wills1976};${}^{10}$ \citep{Donato2001};${}^{11}$\citep{Ghisellini2011};${}^{12}$\citep{Nieppola2005};}\\
\multicolumn{10}{l}{${}^{13}$\citep{Fan2012};${}^{14}$\url{http://tevcat.uchicago.edu}}
\end{tabular}
\end{table*}

The data presented in this paper were taken with the $R$-band filter. Magnitudes were calculated using calibrated field
stars either found in the literature or presented in PTF (Palomar Transient Factory) $R$-band catalogue \citep{Ofek2012}
or USNO-B1.0 catalogue \citep{Monet2003}, depending on availability.

The exposure length was adjusted by the brightness of each target, which was estimated during the short pointing exposures,
depending also on the sky conditions. The average photometric error in magnitudes is $0.04$ mag. The data were processed using the
specialized pipeline described in detail by \cite{King2014} along with the telescope control system.

Since we have introduced a Galactic latitude cut selecting objects with $|b| > 10\dg$, the average colour excess in the directions of our
targets is relatively low, $\overline{E(B-V)} = 0.11^m$ \citep{Schlafly2011}, implying that the interstellar polarization
is less than $1.0\%$ on average \citep{Serkowski1975}. The statistical uncertainty in the degree of polarization is less
than $1\%$ in most cases, while the EVPA is typically determined with a precision of $1^{\circ}$ -- $10^{\circ}$ depending on the source
brightness and fractional polarization. Detailed description of the instrument model and error analysis is given in \cite{King2014}.

In order to resolve the $180\dg$ ambiguity of the EVPA we followed a standard procedure
\citep[see e.g.][]{Abdo2010,Ikejiri2011,Kiehlmann2013}, which is based on the assumption that temporal variations of the
EVPA are smooth and gradual, hence adopting minimal changes of the EVPA between consecutive measurements. We define the EVPA
variation as $\Delta \theta_{\rm n} = |\theta_{{\rm n}+1} - \theta_{\rm n}| - \sqrt{\sigma(\theta_{{\rm n}+1})^2 + \sigma(\theta_{\rm n})^2} $, where
$\theta_{{\rm n}+1}$ and $\theta_{\rm n}$ are the $n+1$ and $n$-th points of the EVPA curve and $\sigma(\theta_{{\rm n}+1})$ and
$\sigma(\theta_{\rm n})$ are the corresponding errors of the position angles. If $\Delta \theta_{\rm n} > 90\dg$, we shift the
angle $\theta_{{\rm n}+1}$ by $\pm \, n \times 180\dg$, where the integer $\pm \, n$ is chosen in such a way that it minimizes
$\Delta \theta_{\rm n}$. If $\Delta \theta_{\rm n} \leq 90\dg$, we leave $\theta_{{\rm n}+1}$ unchanged.

Our first period of regular photometric and polarimetric monitoring of blazars started in 2013 July and lasted until the
end of 2013 November. During the five-month period we obtained more than 1100 measurements of 101 objects from our sample
almost uniformly spread over the observing season of each object. The median cadence and total season length for objects
with $\Delta t$-median smaller than 20 d (including the June survey data, \citealp{Pavlidou2014}) is presented in
Fig.~\ref{fig:cadence_length}, which is discussed in more detail in Sec.~\ref{subsec:rot_freq}.

\subsection{Gamma-ray observations} \label{subsec:gamma_obs}

The gamma-ray data were obtained with the Large Area Telescope (LAT) onboard the {\em Fermi} gamma-ray space observatory,
which observes the entire sky every 3 h at energies of 20 MeV -- 300 GeV \citep{Atwood2009}. We analysed LAT data in
the energy range $100\, {\rm MeV} \le E \le 100\, {\rm GeV}$ using the unbinned likelihood analysis of the standard
{\em Fermi} analysis software package Science Tools v9r33p0 and the instrument response function $P7REP\_SOURCE\_V15$. Source
class photons (evclass=2) were selected within a $15\dg$ region of interest centred on a blazar. Cuts on the satellite
zenith angle ($< 100^{\circ}$) and rocking angle ($< 52^{\circ}$) were used to exclude the Earth limb background. The
diffuse emission from the Galaxy was modelled using the spatial model $gll\_iem\_v05\_rev1$. The extragalactic diffuse and
residual instrumental backgrounds were included in the fit as an isotropic spectral template $iso\_source\_v05$. The
background models\footnote{\url{http://fermi.gsfc.nasa.gov/ssc/data/access/lat/2yr_catalog/gll_psc_v07.xml}} include all
sources from the 2FGL catalogue within $15\dg$ of the blazar. Photon fluxes of sources beyond $10\dg$ from the blazar
and spectral shapes of all targets were fixed to their values reported in 2FGL. The source is considered to be detected
if the test statistic, TS, provided by the analysis exceeds 10, which corresponds to approximately a $3\sigma$ detection
level \citep{Nolan2012}. The systematic uncertainties in the effective LAT area do not exceed 10 per cent
in the energy range we use \citep{Ackermann2012}. This makes them insignificant with respect to the statistical
errors, that dominate over the short time-scales analysed in this paper. Moreover our analysis is based on the relative
flux variations. Therefore the systematic uncertainties were not taken into account.

Different time bins $t_{\rm int}$, from 1 week to 25 d were used, depending on the flux density of the object. In order
to make the analysis more robust we increased sampling of the photon flux curves shifting centres of the time bins by
$t_{\rm int}/4$ interval from each other. This prevents losses of possible short-term events in the light curves and reduces
the dependence of results on the particular position of the time bins. The oversampling introduces an autocorrelation in
the photon flux curves, which is however inessential for the analysis used in this work.

\begin{figure*}
 \centering
 \includegraphics[width=0.42\textwidth]{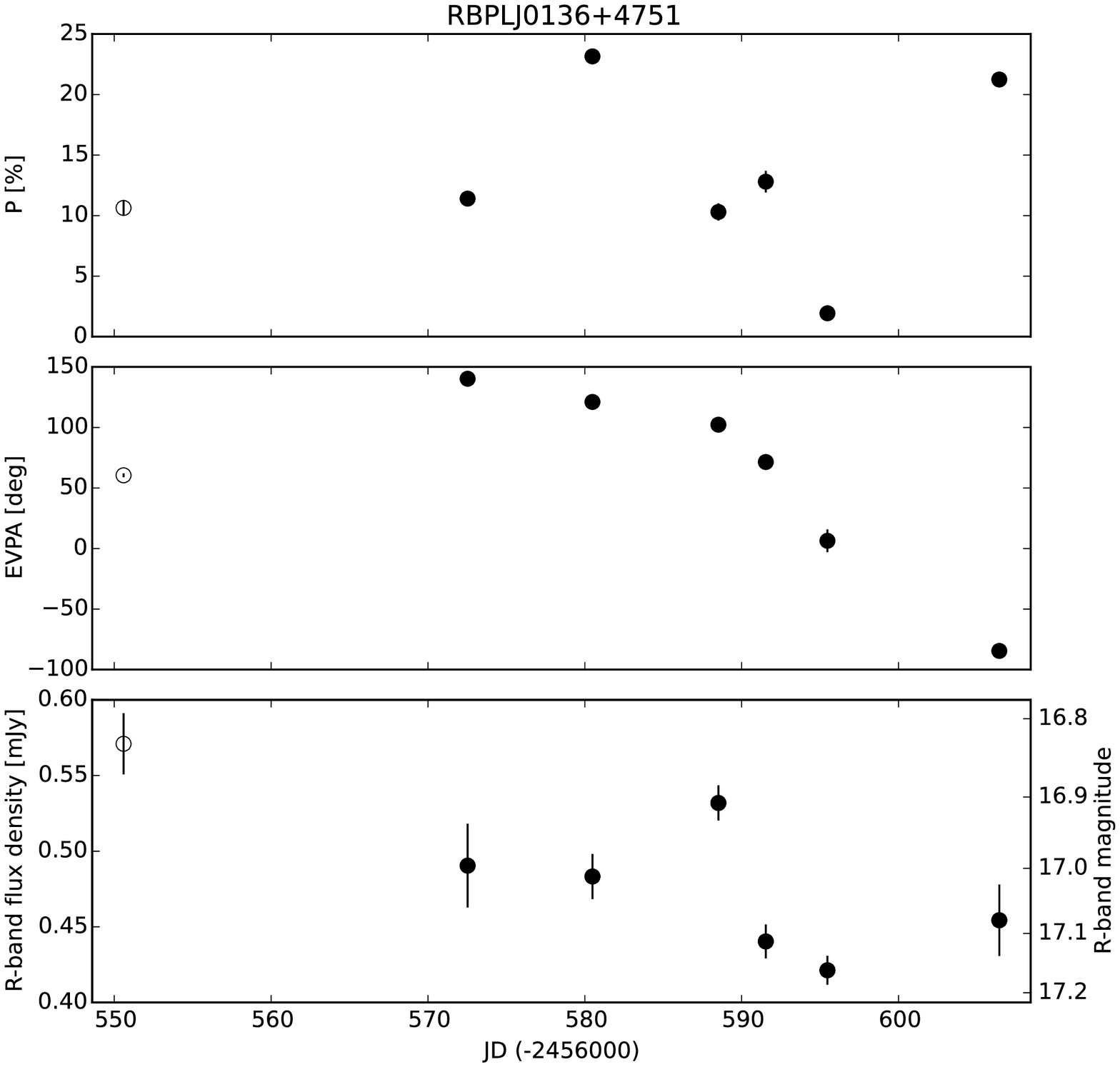}
 \includegraphics[width=0.42\textwidth]{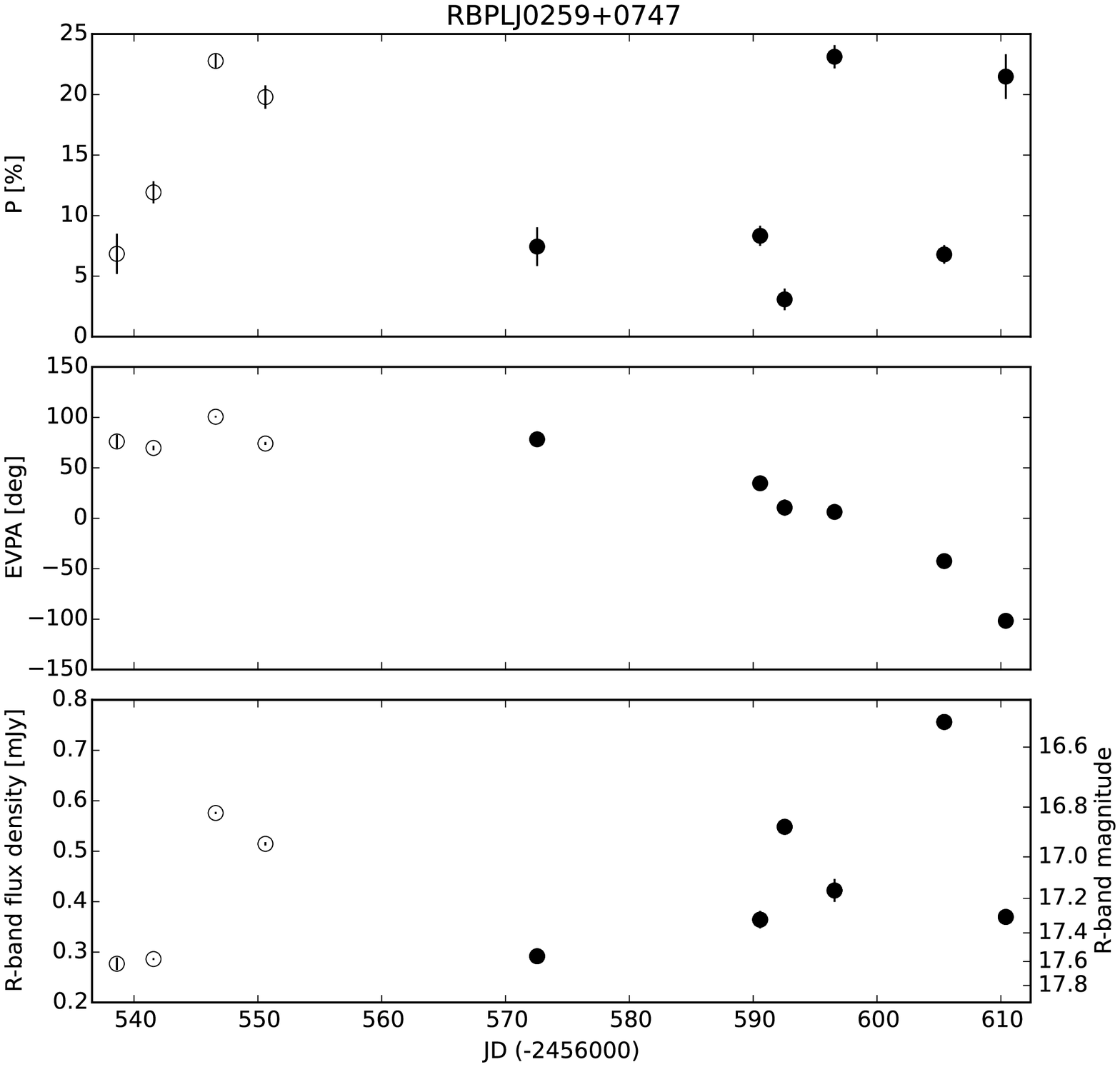}\\
 \includegraphics[width=0.42\textwidth]{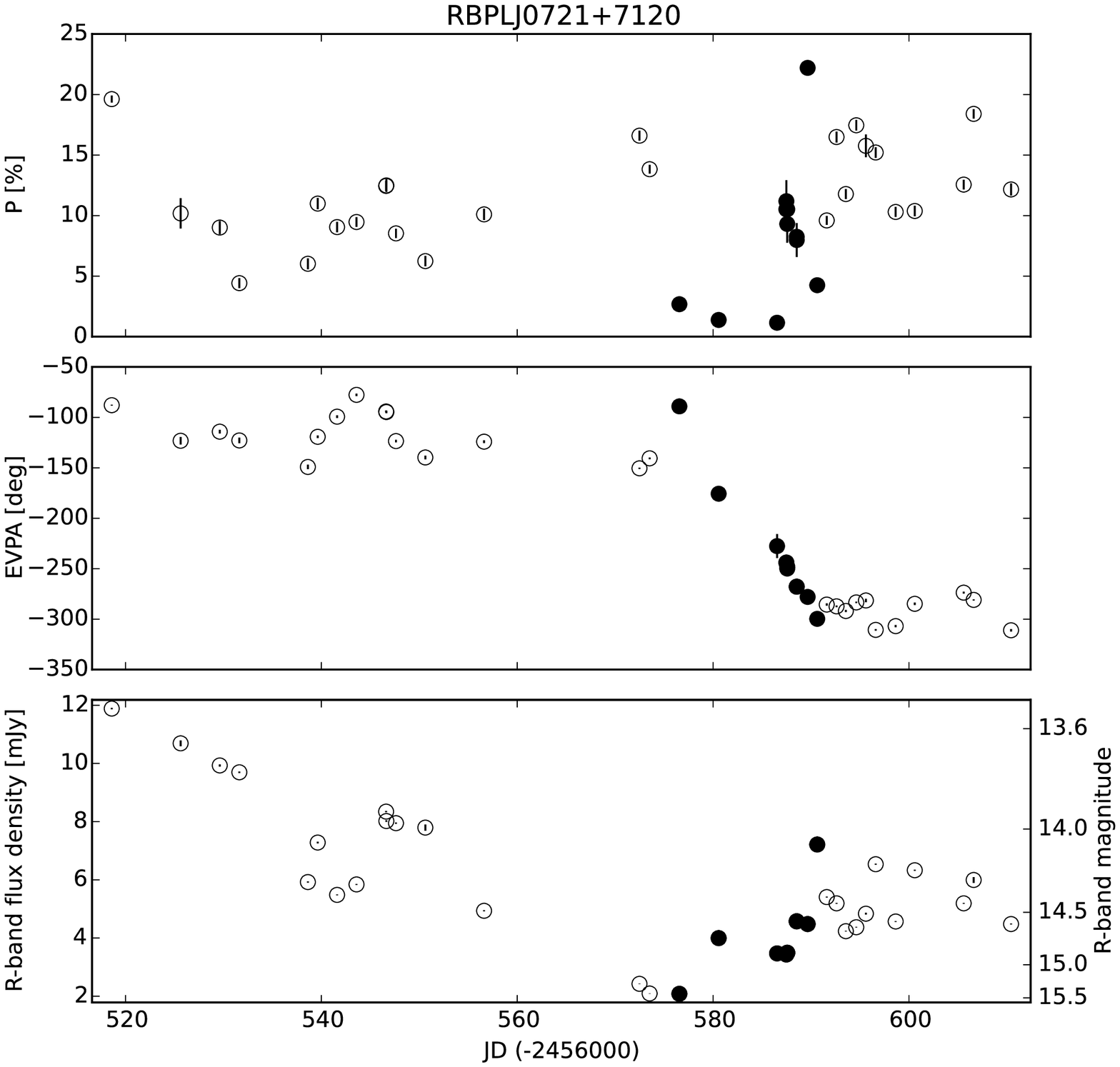}
 \includegraphics[width=0.42\textwidth]{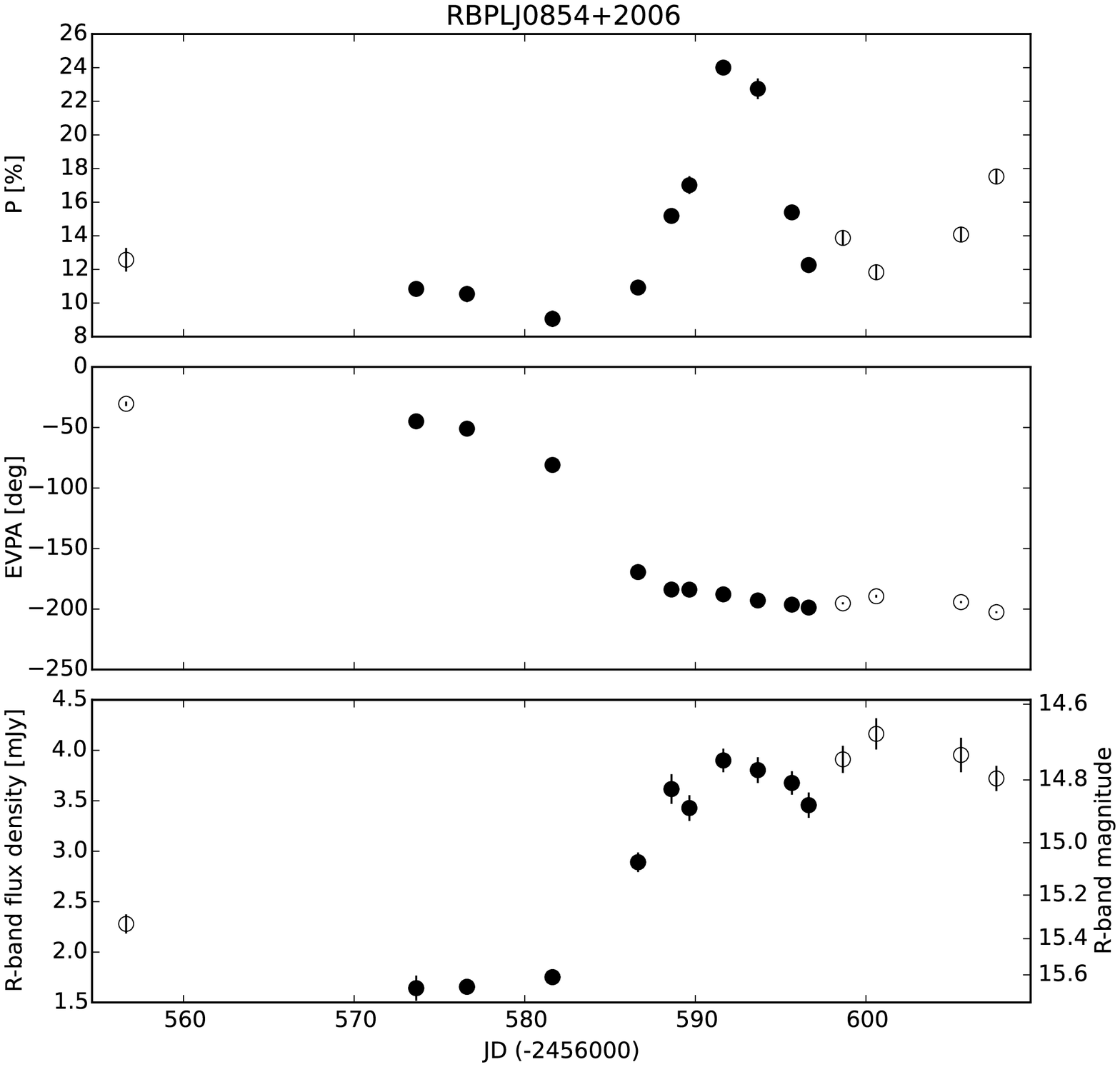}\\
 \includegraphics[width=0.42\textwidth]{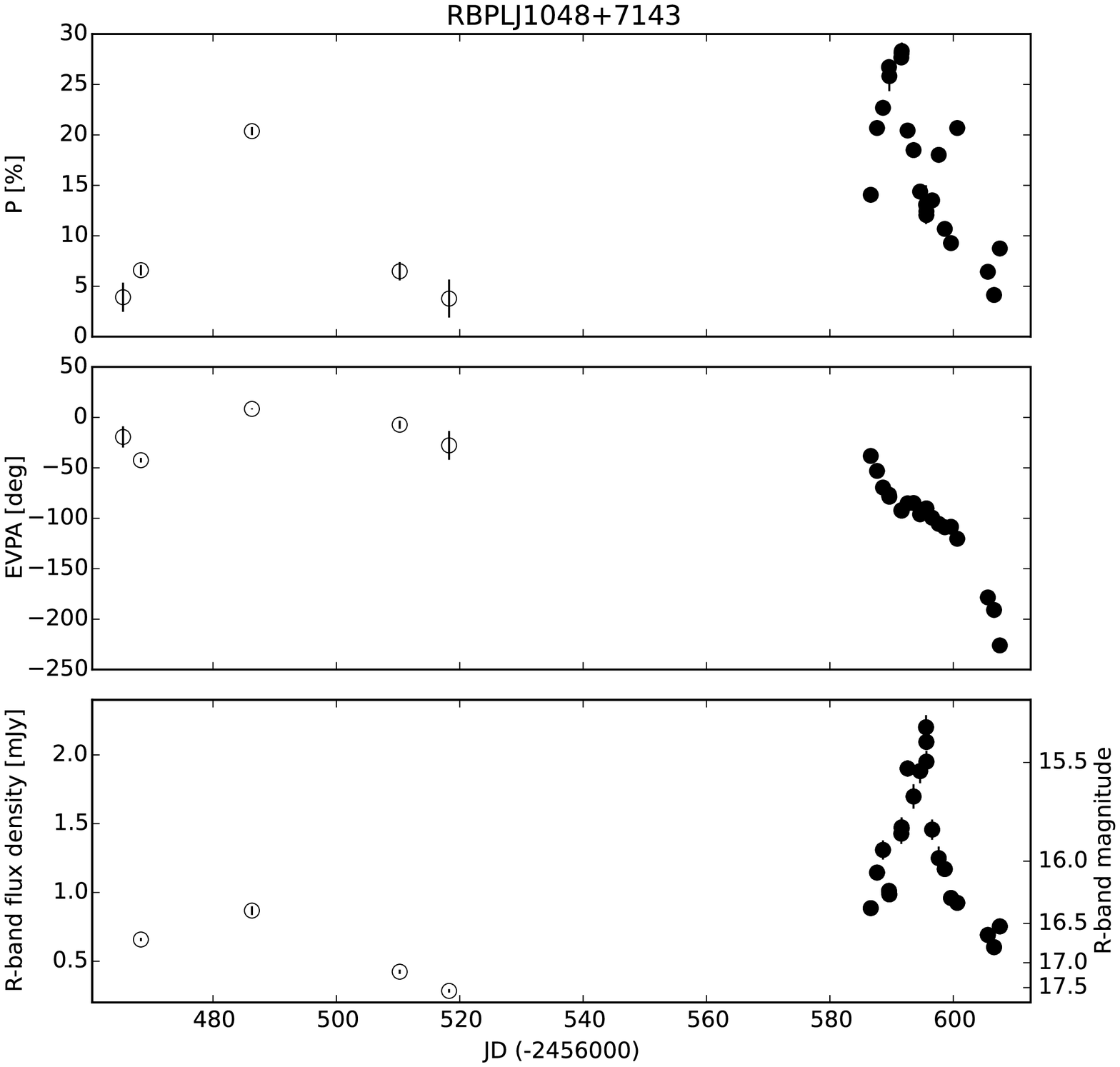}
 \includegraphics[width=0.42\textwidth]{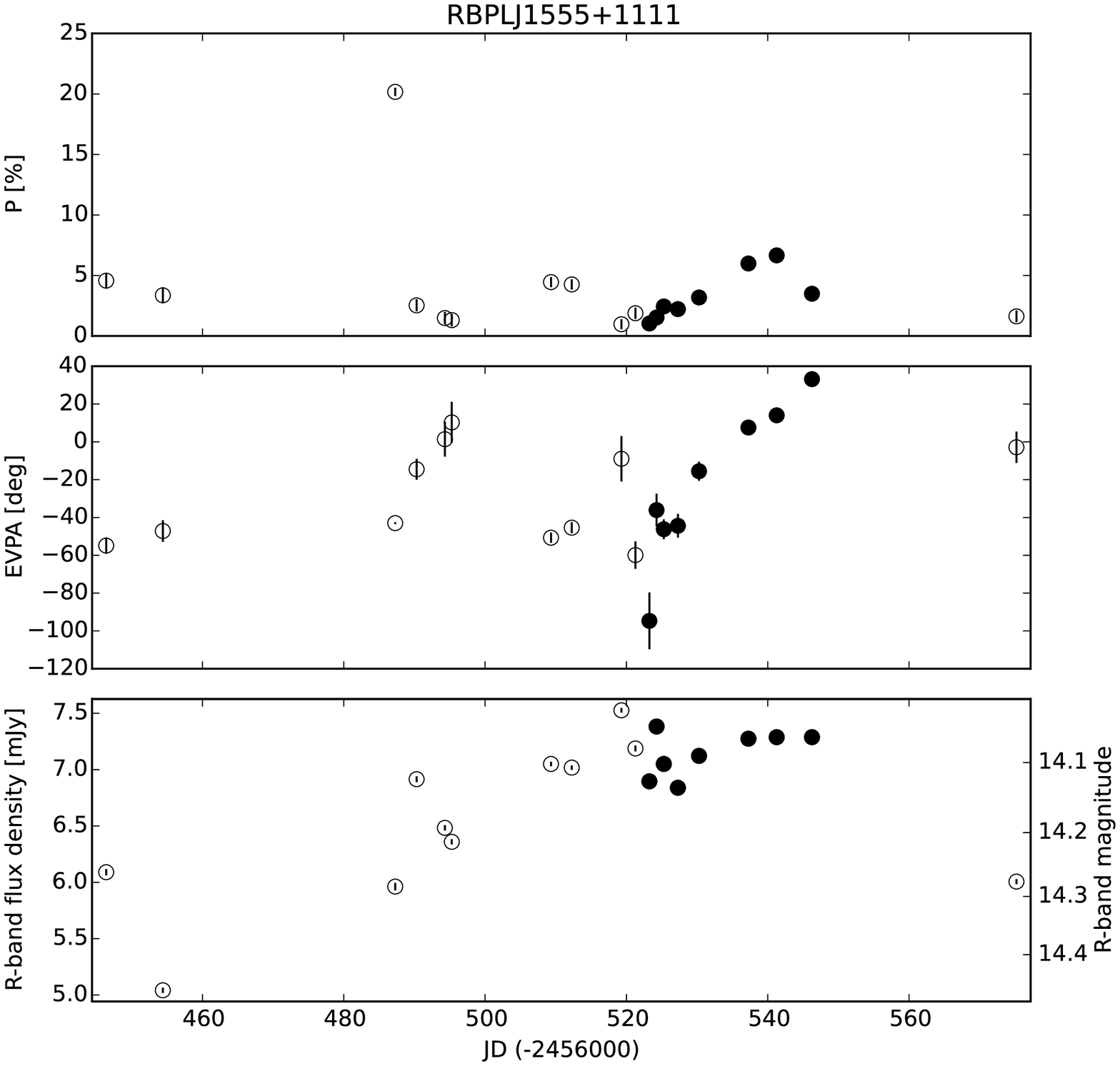}\\
\caption{Evolution of polarization degree, polarization position angle and $R$-band magnitude for blazars with a detected
rotation in the first RoboPol season. Periods of rotations are marked by filled black points.}
\label{fig:rotations}
\end{figure*}

\begin{figure*}
 \centering
 \includegraphics[width=0.42\textwidth]{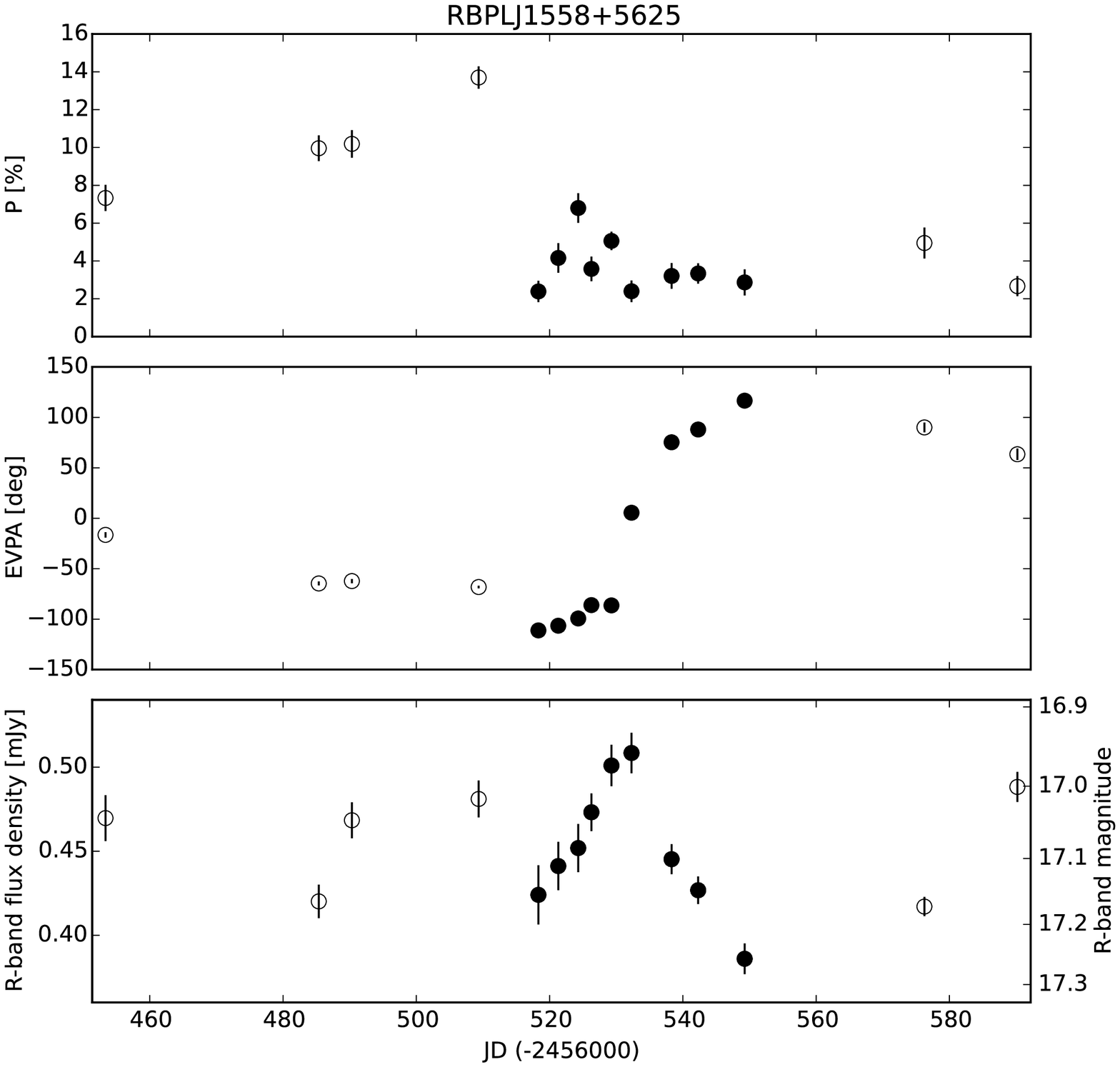}
 \includegraphics[width=0.42\textwidth]{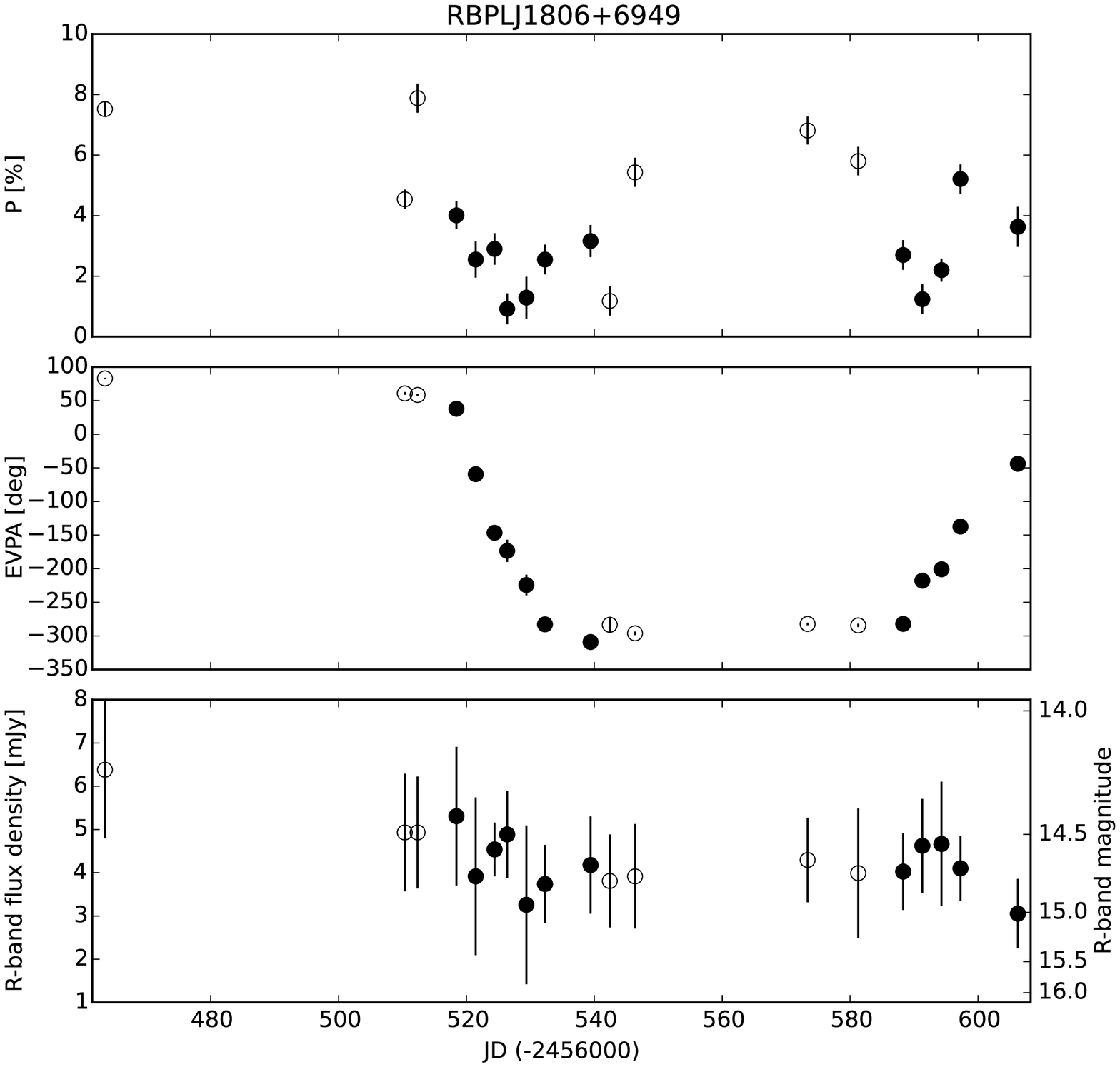}\\
 \includegraphics[width=0.42\textwidth]{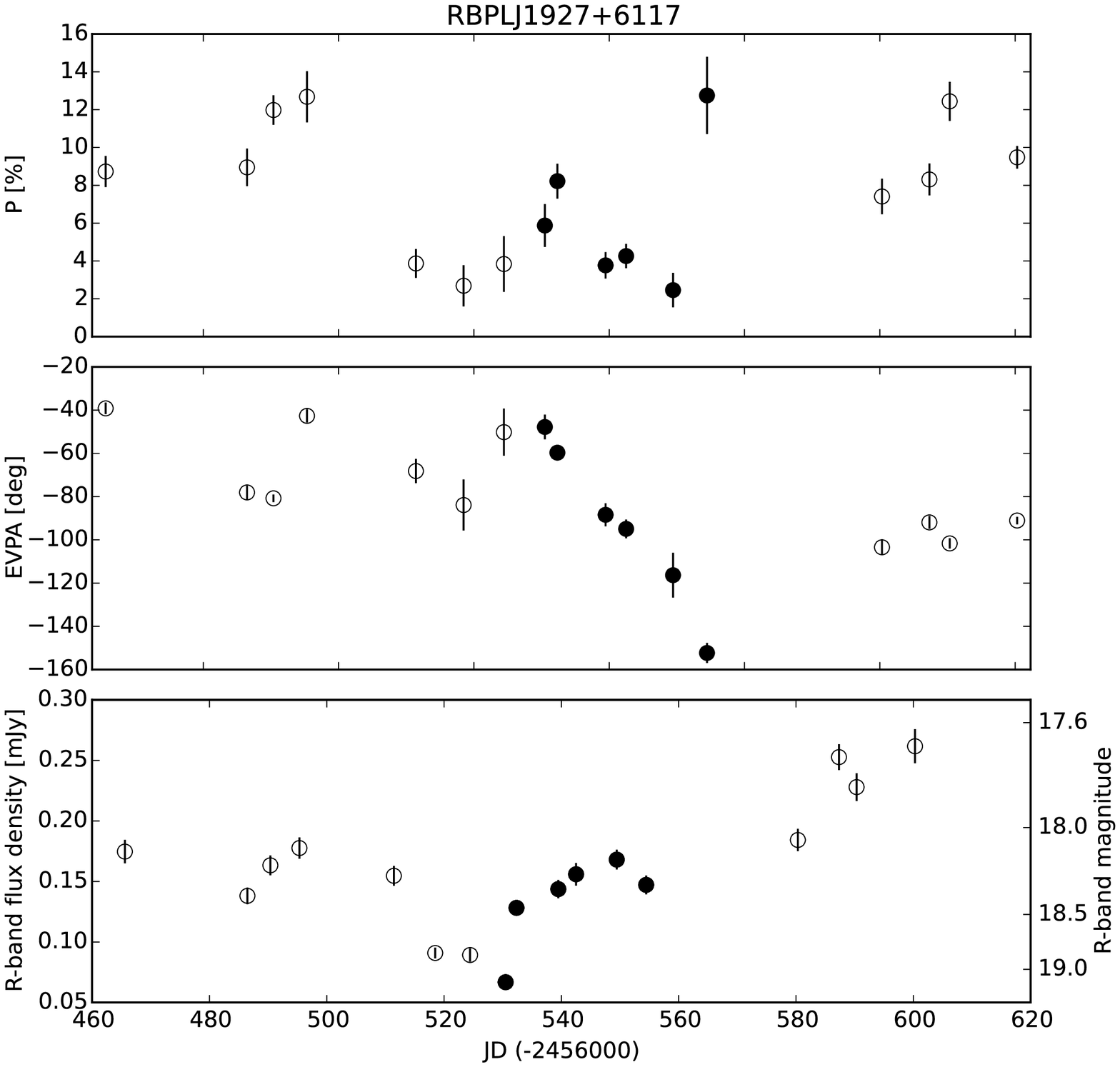}
 \includegraphics[width=0.42\textwidth]{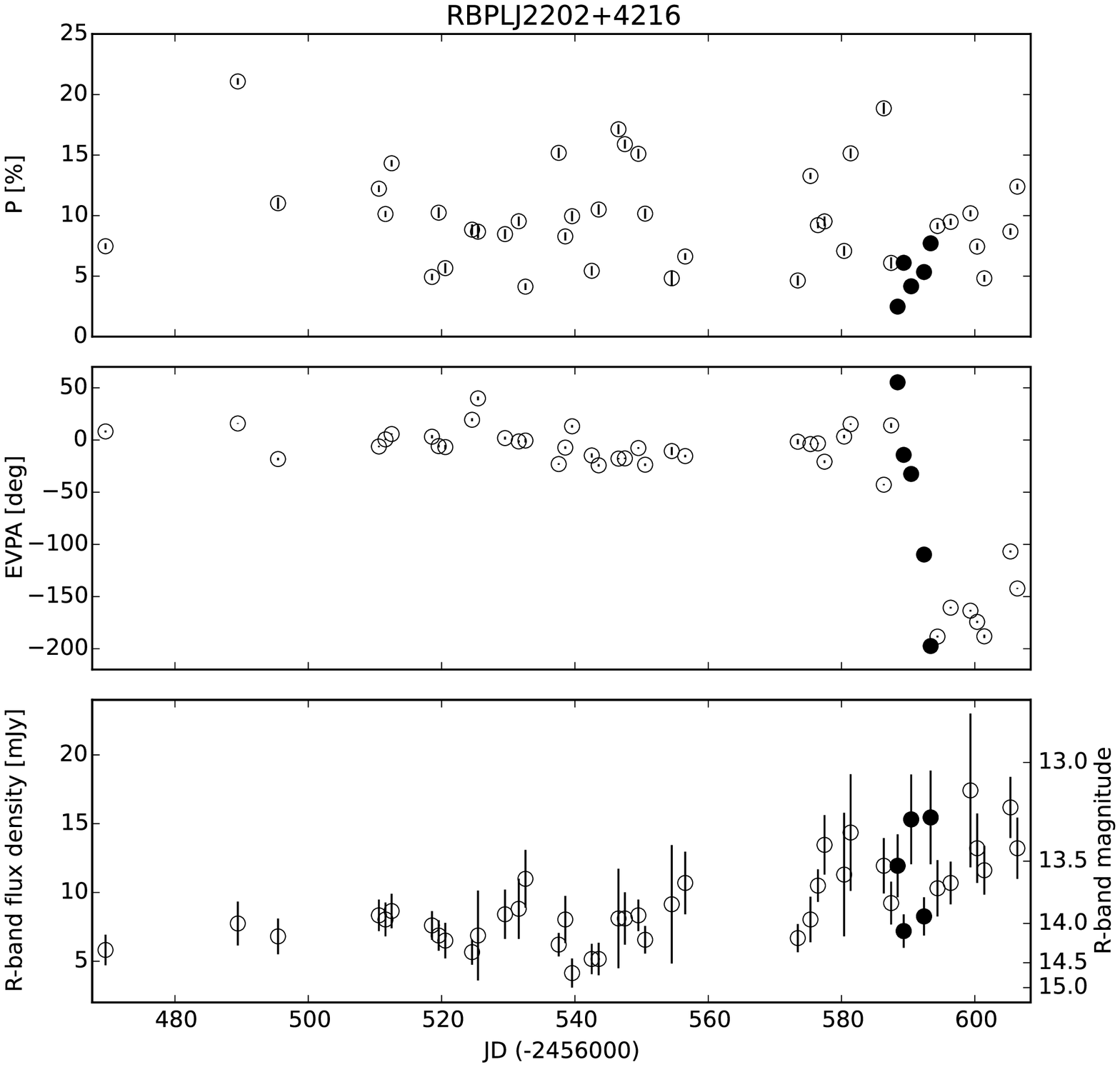}\\
 \includegraphics[width=0.42\textwidth]{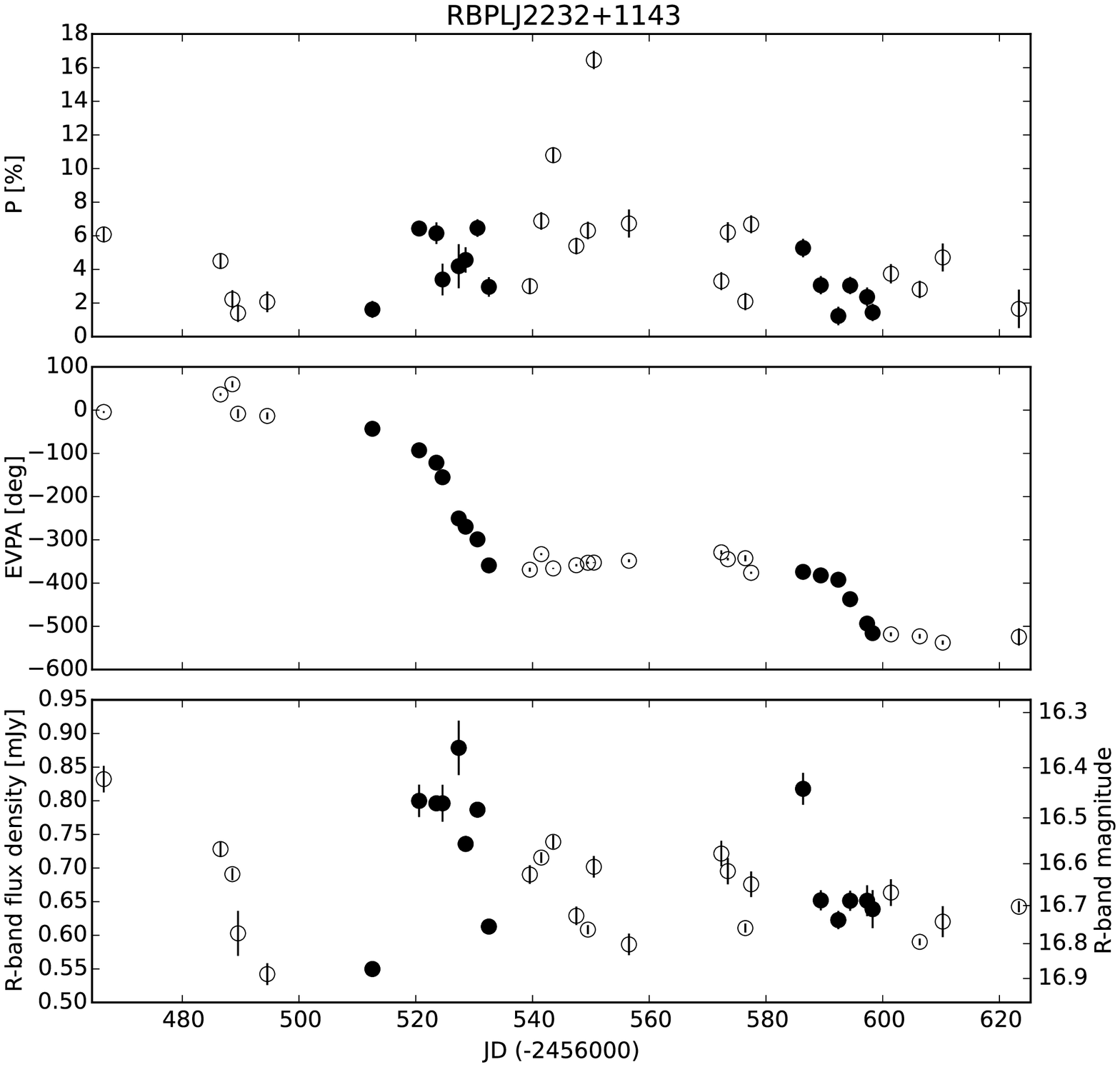}
 \includegraphics[width=0.42\textwidth]{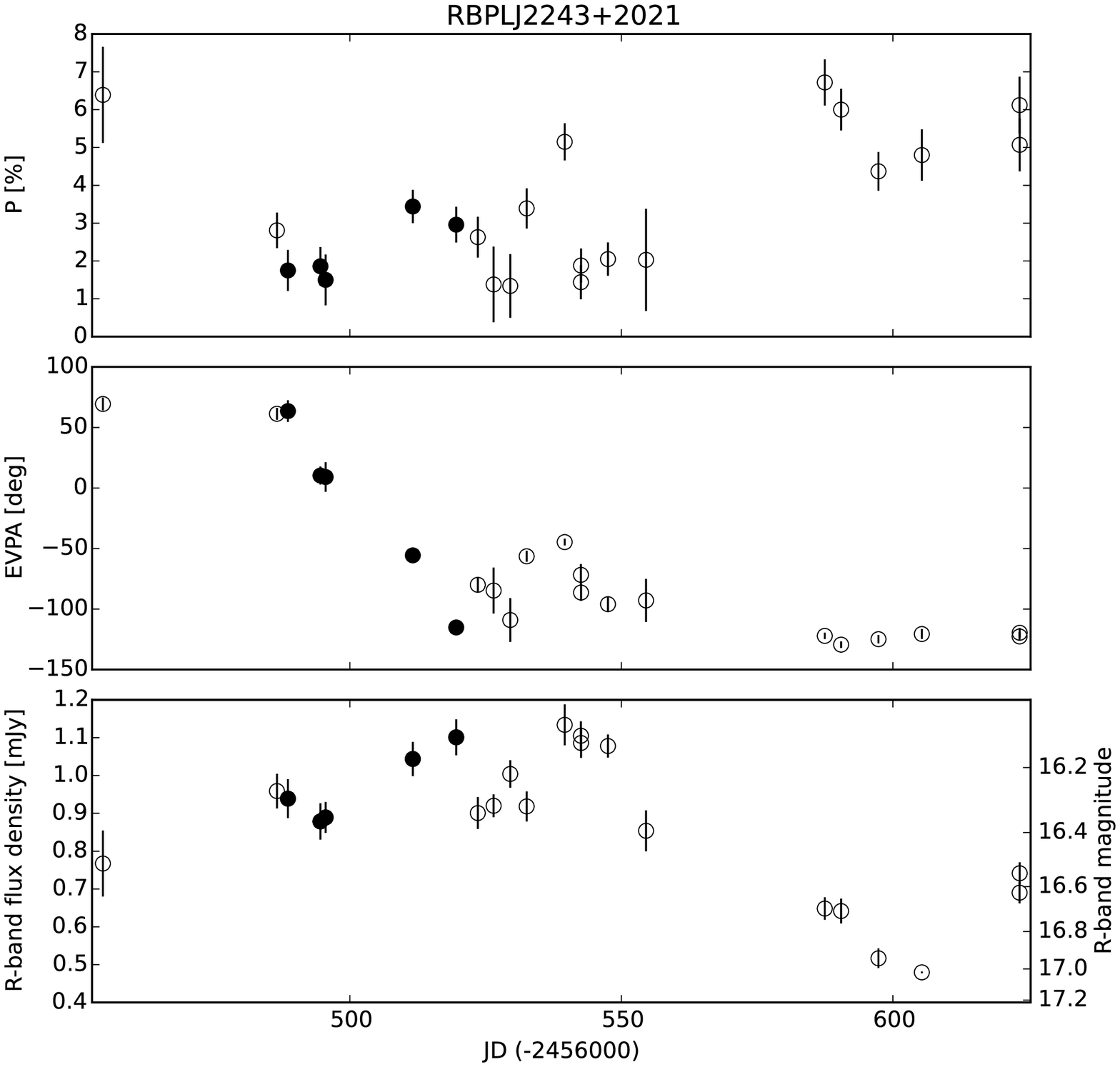}\\
\contcaption{(Continued) Evolution of polarization degree, polarization position angle and $R$-band magnitude for blazars
with a detected rotation in the first RoboPol season. Periods of rotations are marked by filled black points.}
\label{fig:rotations2}
\end{figure*}

\begin{figure*}
 \centering
 \includegraphics[width=0.42\textwidth]{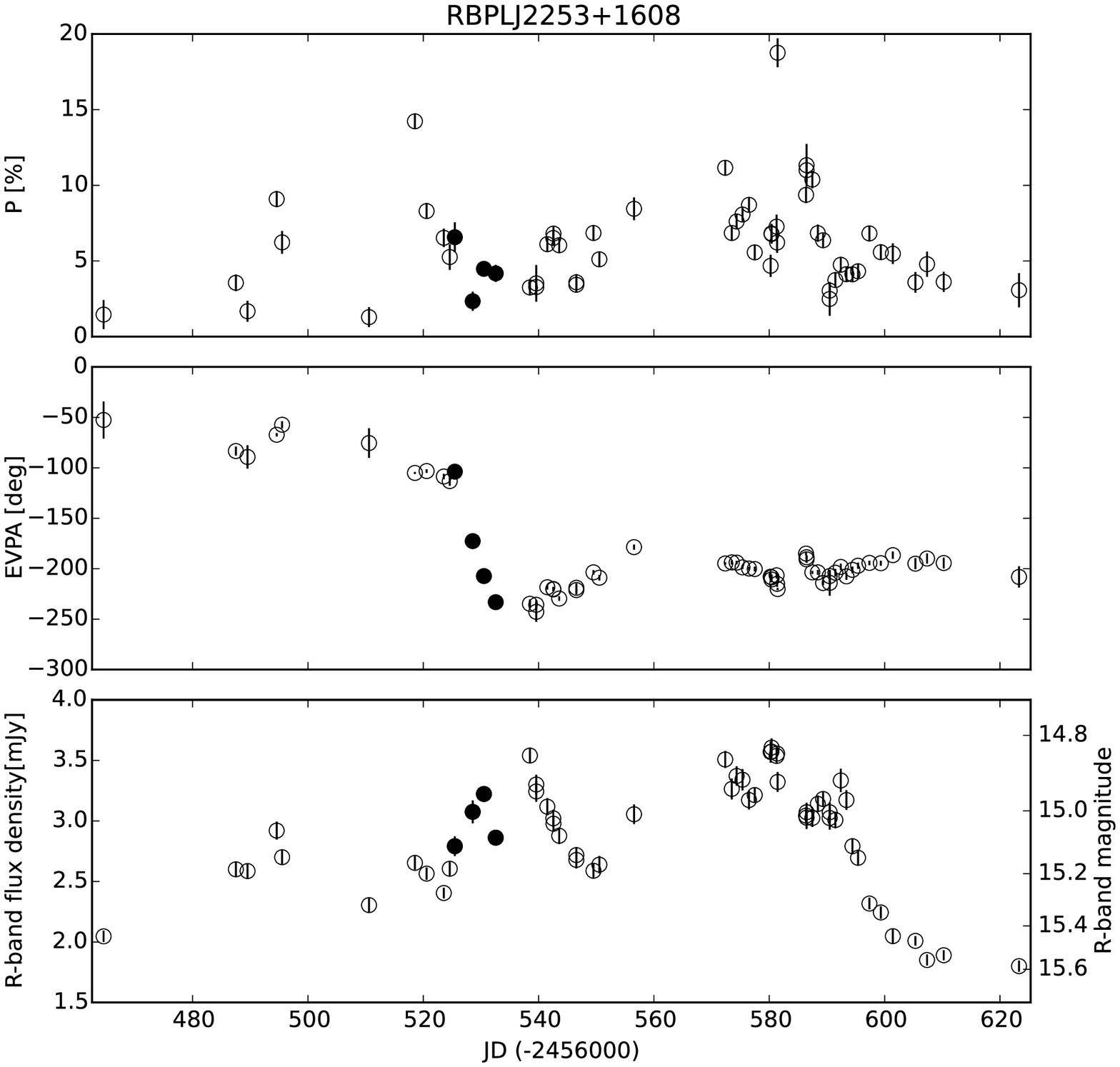}
 \includegraphics[width=0.42\textwidth]{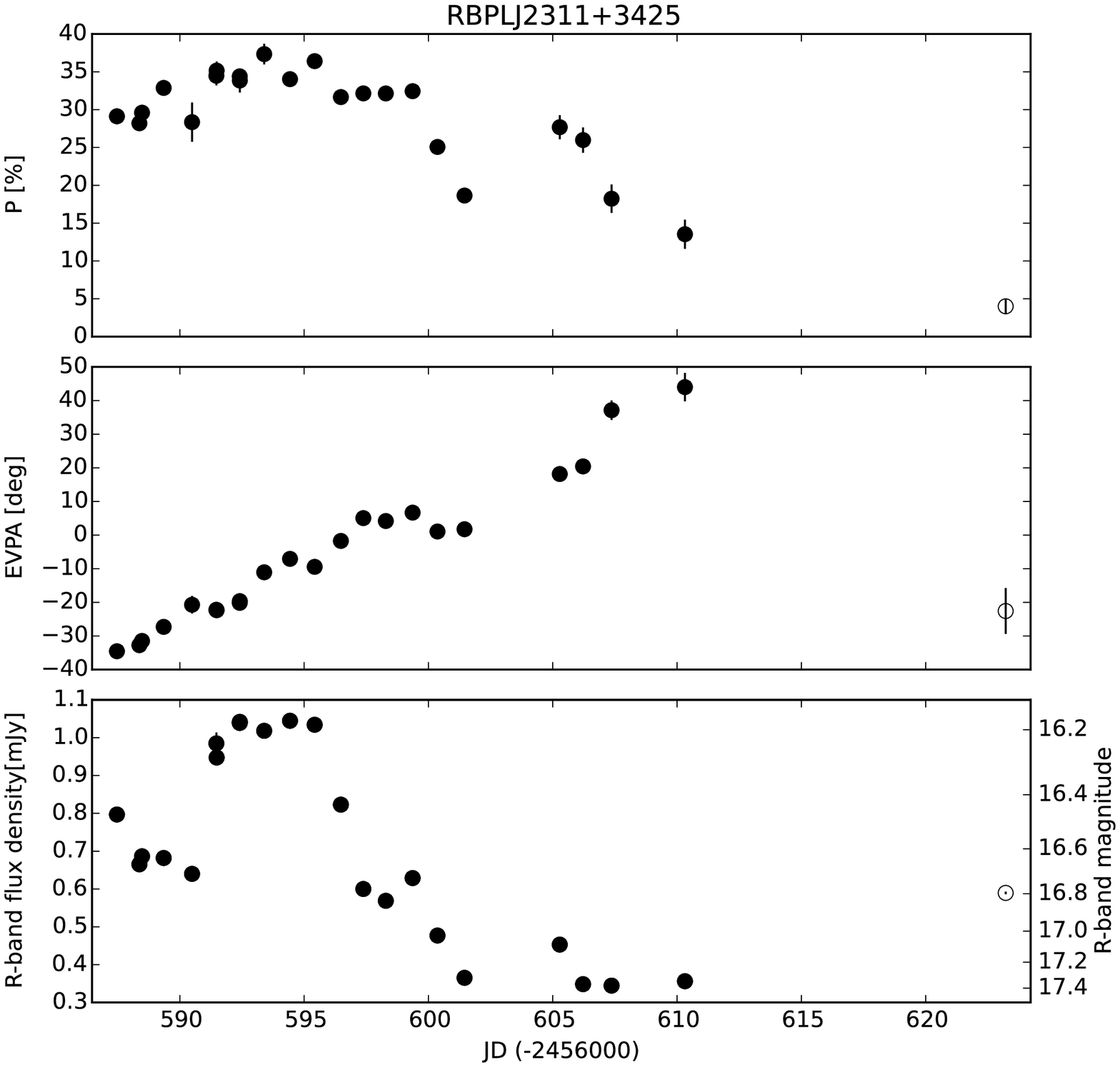}\\
\contcaption{(Continued) Evolution of polarization degree, polarization position angle and $R$-band magnitude for blazars
with a detected rotation in the first RoboPol season. Periods of rotations are marked by filled black points.}
\label{fig:rotations3}
\end{figure*}

\section{Results} \label{sec:rotations}
\subsection{Detected rotations of EVPA} \label{subsec:det_rot}
The optical emission polarization plane of blazars is often variable even within the course of a single night. There is
no objective physical definition of an EVPA rotation. Strictly speaking, any change of the EVPA between two measurements
constitutes a rotation. However typically only high-amplitude ($> 90\dg$), smooth and well tracked variations of the
EVPA are considered as rotations in the literature.

We accept a swing between two consecutive EVPA measurements $\Delta \theta = |\theta_{{\rm i}+1} - \theta_{\rm i}|$ as significant if
$\Delta \theta > \sqrt{\sigma(\theta_{{\rm i}+1})^2 + \sigma(\theta_{\rm i})^2}$. We define as an EVPA rotation any continuous change
of the EVPA curve with a total amplitude $\Delta \theta_{\rm max} > 90\dg$, which is comprised by at least four measurements
with significant swings between them. Start and end points of a rotation event are defined by a change of the EVPA curve
slope $\Delta \theta_{\rm i}/\Delta t_{\rm i}$ by a factor of 5 or a change of its sign. This definition is rather conservative,
and is in general consistent with rotations reported in the literature.

Using this definition, we identified 14 rotations of the EVPA in 12 blazars from the main sample during the season of 2013 
(see Table~\ref{tab:rbpl_rotations}). This number is comparable to the number of previously known events of this type.
Two more blazars with detected rotations, namely RBPLJ0721+7120 and RBPLJ0854+2006, belong to the additional sample of
hand-picked sources. These blazars/events were not included in the statistical or frequency analysis of the following
sections in this paper. The full season EVPA curves along with the evolution of the polarization degree and the $R$-band
flux density, for all 14 blazars with detected rotations, are shown in Fig.~\ref{fig:rotations} and listed in
Table~\ref{tab:rbpl_rotations}. The EVPA rotations are marked by filled black points. Clearly the events we have considered
as rotations based on our criteria are the largest $\Delta \theta_{\rm max}$ rotation events that appear in these data sets.
They are all characterized by smooth variations with a well-defined trend. Two events plotted in Fig.~\ref{fig:rotations}
do not follow the definition strictly. These are the rotation events detected in the data sets of RBPLJ1048+7143 and
RBPLJ2311+3425. In both cases the rotations were interrupted by short, low amplitude, albeit significant swings in the
opposite direction with respect to the overall rotation. Since both events are well sampled these small deviations do not
introduce any significant difference in the overall EVPA trend. Hence both events can be considered as single, large
$\Delta \theta_{\rm max}$ rotations. In addition the RBPLJ2311+3425 event has an amplitude of $\sim 74\dg$, which is
less than the lower limit we accepted. However the start and end points of the rotation are not defined due to a sparse
sampling. It is likely that this well defined EVPA change would meet the $90^{\circ}$ limit if we had a longer data set
for this object. It is for this reason that we include this event in our sample of rotations. Both events have not been
used in any of our statistical analyses involving comparison between simulated and observed rotations.

Some of the EVPA rotation events are coincident with an increase in the total flux, as it follows from a visual inspection
of Fig.~\ref{fig:rotations}. A quantitative comparison between the optical flux and the polarization variations will be
presented in a forthcoming paper.

\begin{table*}
 \centering
 \begin{minipage}{140mm}
  \caption{\label{tab:oldrotations} Data on known rotation of optical EVPA in blazars. Columns (1),(2) - blazar identifiers;
(3) - average rotation rate; (4) - total amplitude of EVPA change; (5) - TeV emission flag according to TeVCat$^{5}$
(``Y'' means that the blazar has been detected in gamma rays with $E>1$ TeV, ``N'' - otherwise); (6) - blazar subclass
(LBL, IBL, HBL denote low, intermediate and high synchrotron peaked BL Lacertae objects, FSRQ -- flat spectrum 
radio quasar); (7) - reference.}
  \label{tab:known_rotations}
  \begin{tabular}{@{}lccccccc@{}}
  \hline
   Blazar ID   &   Survey       & $\langle\frac{\Delta \theta}{\Delta t}\rangle$ & $\Delta \theta_{\rm max}$ & TeV & Class & Reference \\
               &   name         &          (deg d$^{-1}$)                        &          (deg)            &     &       &           \\
 \hline
% data on classes is from  Ciaramella et al. 2004
RBPLJ0423$-$0120 & PKS 0420$-$014 & $-$11.1   & $-$110 &    N    & FSRQ${}^4$   & \citep{DArcangelo2007} \\
RBPLJ0721+7120   & S5 0716$+$71   &   +130    &   +180 &    Y    & LBL${}^1$    & \citep{Larionov2013}   \\
RBPLJ0854+2006   & OJ 287         & $-$17     & $-$180 &    N    & LBL${}^1$    & \citep{Kikuchi1988}    \\
RBPLJ0958+6533   & S4 0954$+$65   &   +18.2   &   +240 &    N    & LBL${}^2$    & \citep{Larionov2011}   \\
RBPLJ1221+2813   & W Comae        &  $\ge+3.0$&   +110 &    Y    & IBL${}^3$    & \citep{Benitez2013}    \\
RBPLJ1256$-$0547 & 3C 279         &  $-$9     & $-$180 &    Y    & FSRQ${}^4$   & \citep{Abdo2010}       \\
RBPLJ1256$-$0547 & 3C 279         &    +4.3   &   +290 & \ditto  & \ditto       & \citep{Larionov2008}    \\
RBPLJ1256$-$0547 & 3C 279         &    +4.7   &   +140 & \ditto  & \ditto       & \citep{Aleksic2014a}    \\
RBPLJ1512$-$0905 & PKS 1510$-$089 &   +15.6   &   +720 &    Y    & FSRQ${}^4$   & \citep{Marscher2010}    \\
RBPLJ1512$-$0905 & PKS 1510$-$089 &   +12     &   +400 & \ditto  & \ditto       & \citep{Aleksic2014}    \\
RBPLJ1512$-$0905 & PKS 1510$-$089 & $-$50     & $-$250 & \ditto  & \ditto       & \citep{Aleksic2014}    \\
RBPLJ1512$-$0905 & PKS 1510$-$089 &   +11.7   &   +500 & \ditto  & \ditto       & \citep{Sasada2011}${}^6$\\
RBPLJ2202+4216   & BL Lac         &   +46     &   +220 &    Y    & IBL${}^1$    & \citep{Marscher2008}    \\
RBPLJ2202+4216   & BL Lac         &   +21     &   +210 & \ditto  & \ditto       & \citep{Sillanpaa1993}    \\
RBPLJ2232+1143   & CTA 102        & $-$60     & $-$180 &    N    & FSRQ${}^4$   & \citep{Larionov2013b}    \\
RBPLJ2253+1608   & 3C 454.3       &   +16.3   &   +130 &    N    & FSRQ${}^4$   & \citep{Sasada2010}      \\
RBPLJ2253+1608   & 3C 454.3       &   +9.3    &   +400 & \ditto  & \ditto       & \citep{Sasada2012}      \\
\hline
\multicolumn{7}{l}{${}^1$\citep{Donato2001};${}^2$\citep{Ghisellini2011};${}^3$\citep{Tagliaferri2000};${}^4$\citep{Fan2012}}\\
\multicolumn{7}{l}{${}^5$\url{http://tevcat.uchicago.edu};${}^6$same as in \cite{Marscher2010}.}
\end{tabular}
\end{minipage}
\end{table*}

\subsection{General properties of EVPA rotations and rotators} \label{subsec:rot_general}

We estimated the maximal amplitude $\Delta \theta_{\rm max}$ and the duration of the rotations $T_{\rm rot}$, using the
first and last points of each event. Due to a moderate sampling and $180^{\circ}$ EVPA ambiguity, the rotation start
and/or end points cannot be pinpointed accurately in five cases (namely RBPLJ0136+4761, RBPLJ0259+0747, RBPLJ1048+7143,
RBPLJ1806+6949 and RBPLJ2311+3425). This ambiguity affects the estimated $\Delta \theta_{\rm max}$ and $T_{\rm rot}$ of the
event, which should really be considered as lower limits in this case. We also estimated the average rotation rate as
$\langle\Delta \theta/\Delta t \rangle = \Delta \theta_{\rm max} / T_{\rm rot}$.  These parameters as well as the blazar
class and the TeV emission flag are listed in Table~\ref{tab:rbpl_rotations}.

\begin{figure}
 \centering
 \includegraphics[width=0.23\textwidth]{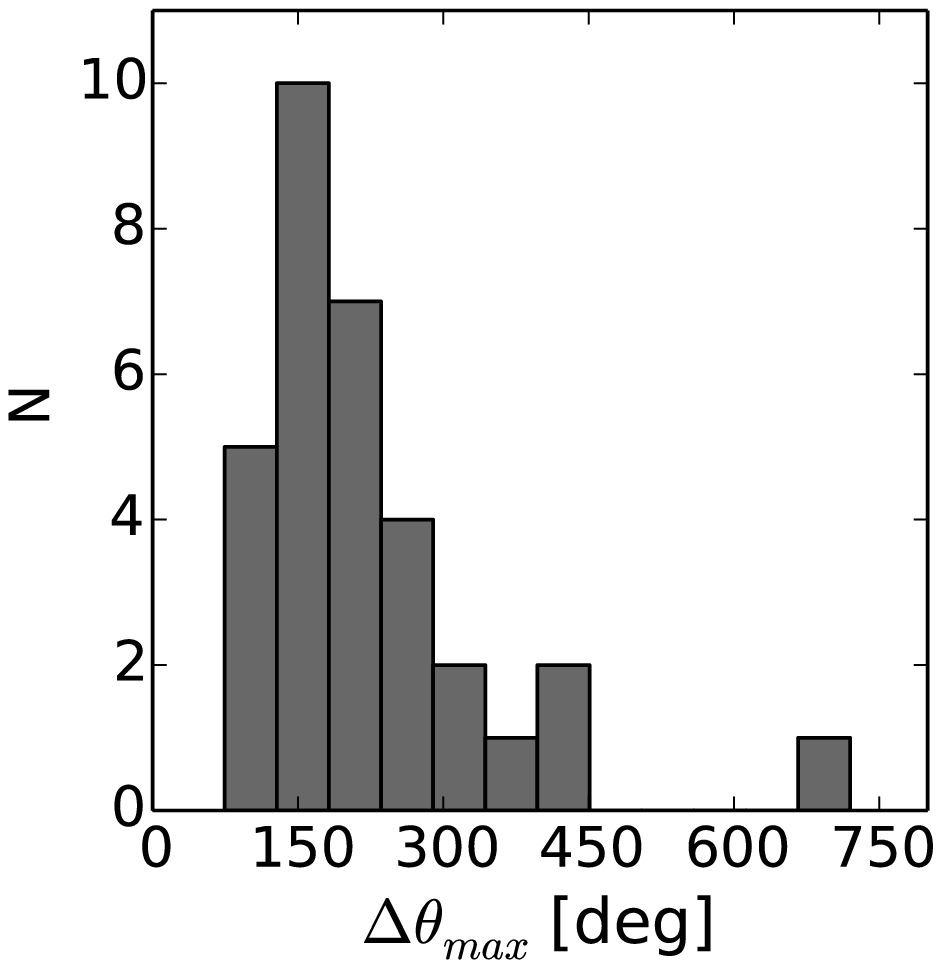}
 \includegraphics[width=0.23\textwidth]{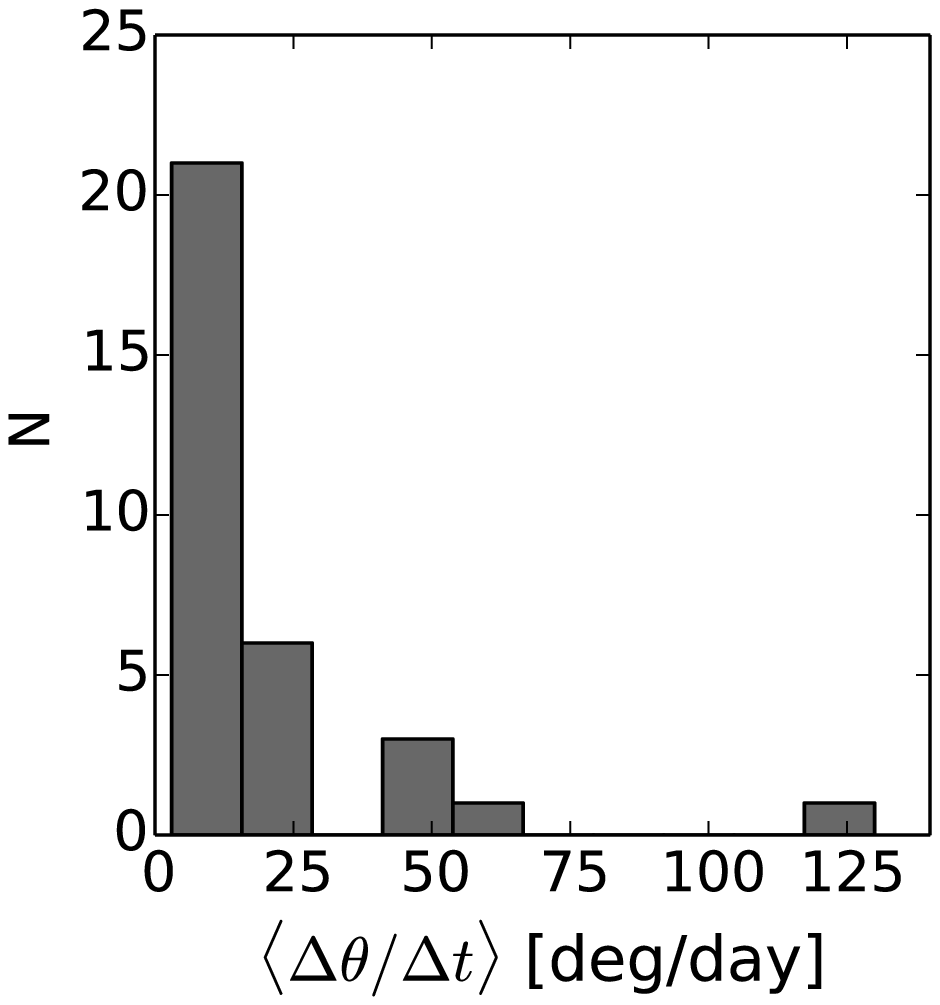}
\caption{Distributions of amplitudes and rates of EVPA rotations detected in RoboPol's first season and reported
in the literature.}
 \label{fig:dist_ampl_rate}
\end{figure}

We also collected data from the literature on previously known rotations of EVPA in blazars which show this behaviour
(``rotators'' hereafter). Rates and $\Delta \theta_{\rm max}$ of these rotations were estimated from plots in the respective
papers. These parameters as well as the blazar class and the TeV emission flag are listed in Table~\ref{tab:known_rotations}.

The distribution of $\Delta \theta_{\rm max}$ and rates of EVPA rotations from historical and RoboPol data are shown in
Fig.~\ref{fig:dist_ampl_rate}. The number of detected rotations clearly decreases with growing $\Delta \theta_{\rm max}$. At
the same time slow rotations dominate in the sample. This is presumably caused by a selection effect, because fast rotations
require better sampling of observations.

Summarizing data on all known EVPA rotations in blazars to date we can list the following properties:
\begin{enumerate}
 \item all known blazars with detected EVPA rotations are in the 2FGL catalogue (i.e. they are ``gamma-ray--loud'' sources);
 \item there are blazars known as TeV emitters as well as non-TeV sources among rotators;
 \item all subclasses of blazars show rotations of the EVPA, regardless of the position of the synchrotron peak maximum
 or the BL Lac/FSRQ dichotomy;
 \item there are eight blazars with more than one rotation detected. Comparison of these rotations shows that a single source
 can show rotations in both directions (five blazars known so far with this behaviour) and rotations observed in the same
 source can be of significantly different rates (in seven blazars rates differ by a factor larger than two in speed).
\end{enumerate}

\subsection{Observed frequency of EVPA rotations} \label{subsec:rot_freq}

The efficiency of an EVPA rotation detection depends on the intrinsic rate of the rotation as well as the frequency and
uniformity of the observing cadence. The ambiguity of the polarization position angle introduces an upper limit on the
rotation rate that can be unequivocally detected with a given typical cadence of observations. Clearly, for a typical
time interval between observations $\langle\Delta t\rangle$, no EVPA rotation with a rate higher than
$90\dg/\langle\Delta t\rangle$ can be observed.

For each blazar in our sample we found the median time difference between successive observations $\Delta t$-median and
the total observing season length (defined as the time difference between the first and the last observations) $T_{\rm obs}$.
These quantities (for blazars observed with $\Delta t$-median $\le 20 $ days) are shown in Fig.~\ref{fig:cadence_length}.
In the same figure, we also plot three lines which indicate the necessary $\Delta t$-median and $T_{\rm obs}$ for detection
of EVPA rotations at rates $\le 10$ (solid line), $\le 15$ (dashed line) and $\le 20$ (dotted line) degrees per day.

The leftmost vertical part of each line represents the shortest $T_{\rm obs}$ needed to detect a rotation of
$\Delta \theta_{\rm max} = 90^{\circ}$ at a given rotation rate. The inclined portion of each line is determined by our
requirement on a rotation event to be comprised by a minimum of four points. Given this requirement, as $\Delta t$-median
increases, so does $T_{\rm obs}$. An EVPA data set with $\Delta t$-median and $T_{\rm obs}$ on that line can allow
detection of EVPA rotations with $90\dg \le \Delta \theta_{\rm max} \le 270\dg$. The horizontal part indicates the
maximum $\Delta t$-median allowed the detection of a rotation event under the requirement of $\Delta \theta \le 90\dg$ in
EVPA between two consecutive points.

\begin{table*}
\centering
\caption{Sources of the main and control samples within the $\langle \frac{\Delta \theta}{\Delta t} \rangle < 10$ deg d$^{-1}$ ``detection box''.}
\label{tab:non_rotators}
  \begin{tabular}{lccclccc}
  \hline
 Blazar ID     &   Survey      & $T_{\rm obs}$  & $\langle\Delta t\rangle$  & Blazar ID     &   Survey      & $T_{\rm obs}$  & $\langle\Delta t\rangle$\\
 (RBPL\dots)   &   name        &   (d)          &  (d)                      & (RBPL\dots)   &   name        &   (d)          &    (d)      \\
\hline
\multicolumn{4}{c}{Main sample}                      & J1838+4802   &  GB6 J1838+4802        &  121  &  7.0 \\
J0045+2127   &  GB6 J0045+2127        &  33   &  4.0 & J1841+3218   &  RXJ1841.7+3218        &  152  &  6.0 \\
J0114+1325   &  GB6 J0114+1325        &  38   &  6.0 & J1903+5540   &  TXS 1902+556          &  135  &  5.0 \\
J0211+1051   &  MG1 J021114+1051      &  85   &  4.0 & J1959+6508   &  1ES 1959+650          &  143  &  5.0 \\
J0217+0837   &  ZS0214+083            &  85   &  6.0 & J2005+7752   &  S5 2007+77            &  140  &  7.0 \\
J0423$-$0120 &  PKS 0420$-$01         &  12   &  4.0 & J2015$-$0137 &  PKS2012$-$017         &  155  &  6.5 \\
J0841+7053   &  4C 71.07              &  71   &  6.0 & J2016$-$0903 &  PMNJ2016$-$0903       &  155  &  7.0 \\
J1512$-$0905 &  PKS 1510$-$08         &  88   &  2.0 & J2022+7611   &  S5 2023+760           &  158  &  7.0 \\
J1542+6129   &  GB6 J1542+6129        &  87   &  4.0 & J2030$-$0622 &  TXS 2027$-$065        &  143  &  5.0 \\
J1553+1256   &  PKS 1551+130          &  132  &  4.0 & J2039$-$1046 &  TXS 2036$-$109        &  144  &  5.5 \\
J1604+5714   &  GB6 J1604+5714        &  135  &  7.0 & J2131$-$0915 &  RBS1752               &  127  &  5.0 \\
J1607+1551   &  4C 15.54              &  136  &  8.0 & J2143+1743   &  OX 169                &  119  &  5.0 \\
J1635+3808   &  4C 38.41              &  121  &  2.0 & J2148+0657   &  4C 6.69               &  152  &  4.5 \\
J1642+3948   &  3C 345                &  148  &  6.0 & J2149+0322   &  PKSB 2147+031         &  169  &  6.5 \\
J1653+3945   &  Mkn 501               &  153  &  4.0 & J2150$-$1410 &  TXS 2147$-$144        &  130  &  8.0 \\
J1725+1152   &  1H 1720+117           &  120  &  3.0 & J2225$-$0457 &  3C 446                &  144  &  6.0 \\
J1748+7005   &  S4 1749+70            &  87   &  3.0 & J2251+4030   &  MG4 J225201+4030      &  177  &  6.5 \\
J1751+0939   &  OT 081                &  154  &  4.0 & J2334+0736   &  TXS 2331+073          &  138  &  8.0 \\
J1754+3212   &  RXJ1754.1+3212        &  134  &  5.0 & J2340+8015   &  BZBJ2340+8015         &  113  &  5.5 \\
J1800+7828   &  S5 1803+784           &  133  &  5.0 & \multicolumn{4}{c}{Control sample}           \\
J1809+2041   &  RXJ1809.3+2041        &  152  &  4.5 & J1551+5806   & SBS1550+582            & 118   & 5.0 \\
J1813+3144   &  B2 1811+31            &  150  &  6.0 & J1638+5720   & S4 1637+57             & 138   & 4.5 \\ 
J1836+3136   &  RXJ1836.2+3136        &  151  &  6.0 & J2042+7508   & 4C +74.26              & 99    & 5.0 \\
\hline
\end{tabular}
\end{table*}

We can now estimate the frequency with which EVPA rotations appear in blazars as follows. Out of the 14 detected rotations
in blazars of the main sample, 8 have rates less than 10 deg d$^{-1}$. There are also 41 main sample (``gamma-ray
bright'') blazars that were observed with $\Delta t$-median and $T_{\rm obs}$ (see Table~\ref{tab:non_rotators}) within
the region defined by the solid line in Fig.~\ref{fig:cadence_length}. The total observing length for these blazars is
6432 d. Thereby we estimate the frequency of ``slow'' rotations (rate $< 10$ deg d$^{-1}$) in the main sample
sources as one rotation in $\sim 800$ days (6432 d / 8 rotations). Following the same reasoning we estimate average
frequencies of rotations for blazars in the main sample with rates $ < 15$ deg d$^{-1}$ and $ < 20$ deg d$^{-1}$
as one rotation in $\sim 490$ d (4912/10) and $\sim 180$ d (2363/13), respectively.

\subsection{EVPA variability in blazars of different samples} \label{subsec:dist_swings}

In order to address the question whether ``the EVPA variability is different in objects where rotations were detected compared
to the rest of the main sample and to the control sample'' we collate all EVPA ``swing'' events and measure their
$\Delta \theta_{\rm max}$ and rates. We define an EVPA ``swing'' as any continuous change of the EVPA curve, without a lower limit in
its $\Delta \theta_{\rm max}$ or in the number of measurements. As before, start and end points of a swing event are defined
by a change of the EVPA curve slope by a factor of 5 or a change of its sign.

\begin{figure}
 \centering
 \includegraphics[width=0.23\textwidth]{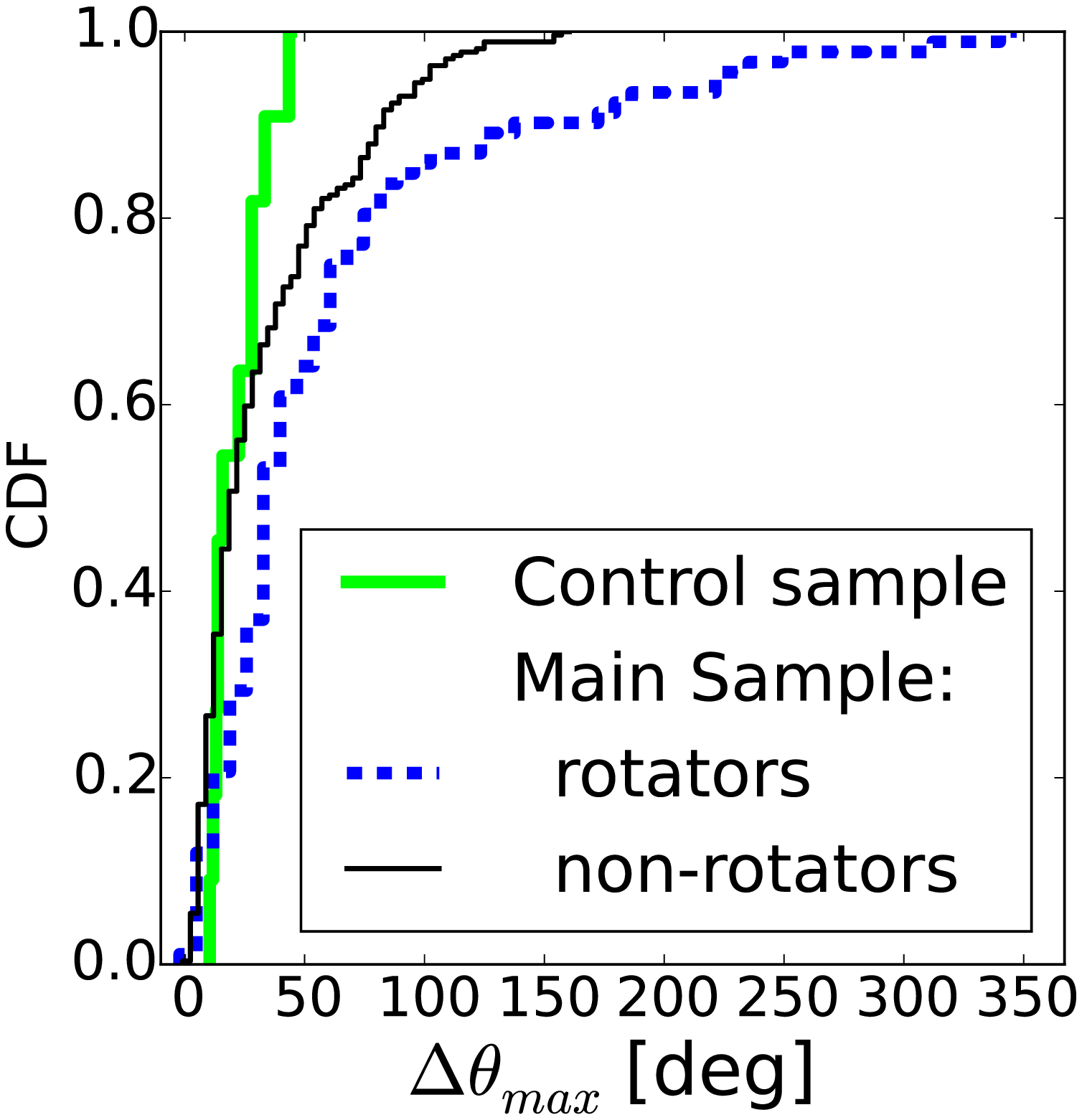}
 \includegraphics[width=0.23\textwidth]{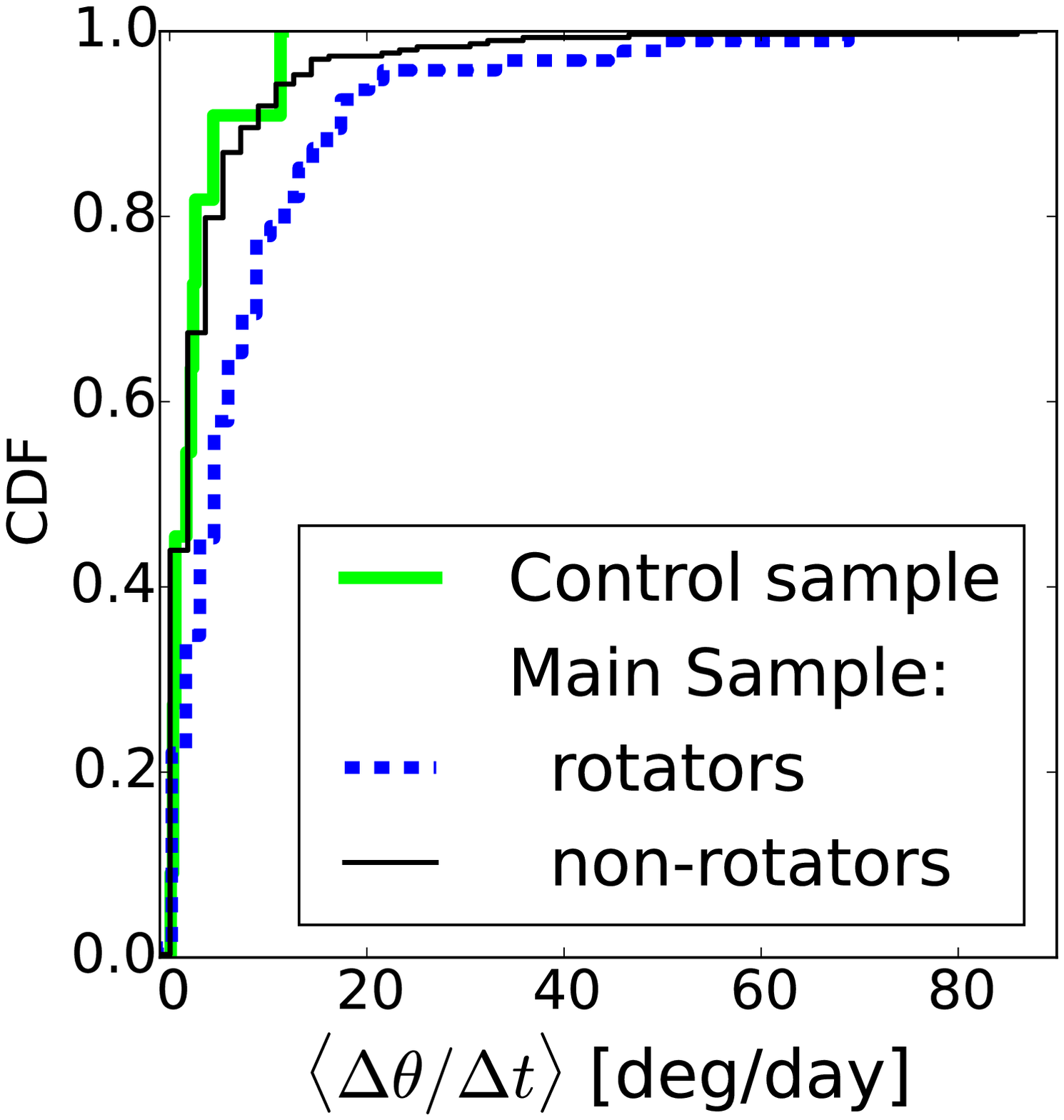}
\caption{CDFs of $\Delta \theta_{\rm max}$ and average rates for the main sample rotators (95 EVPA swings),
the main sample non-rotators (298 swings) and control sample sources (11 swings). See Sec.~\ref{subsec:dist_swings} for details.}
 \label{fig:dist_swings}
\end{figure}

We identified all such events for all blazars of the main and control samples within the 10 deg d$^{-1}$
``detection box'' in Fig.~\ref{fig:cadence_length}, and measured their amplitude, $\Delta \theta_{\rm max}$, and mean
rotation rate. The cumulative distribution function \citep[hereafter CDF; e.g.][]{Wall2012} of the EVPA swings $\Delta \theta_{\rm max}$
and rotation rates for blazars in the main sample which showed rotations (``rotators''), blazars in the main sample, which
did not show rotations (``non-rotators''), as well as for blazars in the control sample, are shown in Fig.~\ref{fig:dist_swings}.

We performed a two sample Kolmogorov--Smirnov (K--S) test \citep[e.g.][]{Wall2012} pairwise for three samples of collected swing amplitudes and rates with
the null-hypothesis that these samples are drawn from the same distribution. The null-hypothesis is rejected for rotators
and non-rotators with the $p\text{-value}=1.2 \times 10^{-5}$, and for rotators and the control sample ($p\text{-value}=5 \times 10^{-3}$).
At the same time the distribution of swing amplitudes in the non-rotators and control sample sources is indistinguishable
according to the test ($p\text{-value}=0.35$). The maximum difference between the CDFs of non-rotators and rotators
is 0.29. It is reached at $\Delta \theta_{\rm max} \approx 25^{\circ}$.
Even if we exclude the 14 rotations (i.e. the largest $\Delta \theta_{\rm max}$ swings) of the main sample blazars,
rotators still remain different from the non-rotators ($p\text{-value}=2 \times 10^{-3}$).

A similar analysis (as the one for $\Delta \theta_{\rm max}$) for the distributions of EVPA swing rates leads to the same conclusion.
The null-hypothesis is rejected for the rotators and the non-rotators ($p\text{-value}=1.4 \times 10^{-6}$) and rotators vs.
the control sample ($p\text{-value}=5 \times 10^{-3}$), while it can not be rejected for the non-rotators and control sample
($p\text{-value}=0.18$).

{\em We therefore conclude that blazars with detected rotations show significantly larger $\Delta \theta_{\rm max}$ and faster EVPA
variations when compared to blazars with no detected rotations.} This difference cannot be attributed to differences in the
sampling properties of the data sets. Therefore, the lack of detection of EVPA rotations in the ``non-rotators'' member
of the main sample, as well as the blazar in the control sample, may have a physical origin. Most of the non-rotators in
the main and control samples may never show an EVPA rotation.

\section{Random walks as the origin of EVPA rotations} \label{sec:random_walks}
\subsection{MC simulations of EVPA swings} \label{subsec:MC_simul}
Potentially EVPA swings can be explained by a stochastic process, which is physically justified by a presence of many
independent cells in the emission region \citep[e.g.][]{Jones1985,DArcangelo2007}. According to this interpretation, the
magnetic field is turbulent and apparent rotations result from a random walk of the full polarization vector direction
as new cells with random magnetic field orientations appear in the emission region \citep{Marscher2014}. In order to
estimate the probabilities that the EVPA rotations we observed with RoboPol are produced by this kind of multicell random
walk process we performed MC simulations of the stochastic variability of the polarization vector on the
QU plane following \citet{Kiehlmann2013}.

For each blazar where an EVPA rotation event was observed, we created $10^4$ artificial light curves, each one with duration
$T_{\rm obs}$. The time steps $\Delta t_{\rm i}$ between consecutive points were drawn from a truncated power-law distribution,
which approximates well the distribution of the time steps in all observed lightcurves. The parameters of this distribution
($\Delta t_{\rm min}$, $\Delta t_{\rm max}$ and the power-law index) were determined by fitting it to the distribution of observed
$\Delta t_{\rm i}$ for each object.

The total flux density $I_{\rm i}$ emitted at each time step $\Delta t_{\rm i}$, was drawn from a log-normal
distribution. Such a distribution approximates reasonably well the distribution of the observed flux densities for all
blazars. The mean and variance of the log-normal distribution was set equal to the sample mean and variance of the
distribution of the flux density of each blazar.

The maximum possible fractional polarization produced by a uniform magnetic field is
$P_{\rm max} = (\alpha + 1)/(\alpha + 5/3) \approx 0.78$ \citep{Pacholczyk1970}. In the case of unresolved emission region
comprising $N$ independent cells with a uniform magnetic field, but randomly oriented among them, the average fractional
polarization is given by the equation \citep{Hughes1991}:
\begin{equation}
 \langle P_{\rm obs} \rangle \approx \frac{P_{\rm max}}{\sqrt{ N }}.
\end{equation}
We used this equation and the observed average polarization fraction, $\langle P_{\rm obs} \rangle$, to estimate the number
of cells, $N$, for each blazar. Each $k$-th cell at $i$-th time step was assigned a flux density $I_{\rm i,k}$ (which was set equal to
$I_{\rm i}/N$ for all cells at each time step) and a set of fractional Stokes parameters $q_{\rm i,k}$ and $u_{\rm i,k}$.
They were found as
\begin{equation}
 \begin{cases}
  q_{\rm i,k} = q_{\rm i,k}^0 \frac{P_{\rm max}}{\sqrt{(q_{\rm i,k}^0)^2 + (u_{\rm i,k}^0)^2}} \\
  u_{\rm i,k} = u_{\rm i,k}^0 \frac{P_{\rm max}}{\sqrt{(q_{\rm i,k}^0)^2 + (u_{\rm i,k}^0)^2}},
 \end{cases}
\end{equation}
where $q_{\rm i,k}^0$ and $u_{\rm i,k}^0$ are two numbers drawn from the standard normal distribution. Thereby the emission
of each cell has polarization fraction $P_{\rm max}$. The sums $Q_{\rm i}=I_i \sum_{{\rm k}=1}^N q_{\rm i,k}$ and
$U_{\rm i}=I_i \sum_{{\rm k}=1}^N u_{\rm i,k}$ determine the total Stokes parameters of the emitting region at each time step.

\begin{table}
\centering
\caption{Random walk modelling results for EVPA rotations detected by RoboPol in 2013. (1) - blazar identifier;
(2) - occurrence of rotations with $\Delta \theta_{\rm max,simul} \ge \Delta \theta_{\rm max,obs}$ estimated from the simulations;
(3) - probability that a rotation produced by the random walk will be observed in $T_{\rm obs}$.}
\label{tab:rbpl_rotations2}
  \begin{tabular}{lcc}
  \hline
   Blazar ID   &  $T_{\rm occ}$ &  P(RW) \\
               &   (d)          &        \\
 \hline
RBPLJ0136+4751 &   505      &  0.11  \\
RBPLJ0259+0747 &   151      &  0.48  \\
RBPLJ0721+7120 &   325      &  0.28  \\
RBPLJ0854+2006 &   142      &  0.36  \\
RBPLJ1048+7143 &   180      &  0.79  \\
RBPLJ1555+1111 &   128      &  1.00  \\
RBPLJ1558+5625 &   266      &  0.51  \\
RBPLJ1806+6949 &   965      &  0.15  \\
RBPLJ1806+6949 &   259      &  0.55  \\
RBPLJ1927+6117 &   137      &  0.98  \\
RBPLJ2202+4216 &   633      &  0.21  \\
RBPLJ2232+1143 &   1557     &  0.09  \\
RBPLJ2232+1143 &   178      &  0.87  \\
RBPLJ2243+2021 &   183      &  0.92  \\
RBPLJ2253+1608 &   184      &  0.86  \\
RBPLJ2311+3425 &   61       &  0.74  \\
\hline
\end{tabular}
\end{table}

At each time step the Stokes parameters of $N_{\rm var}(\Delta t_{\rm i})$ cells, selected randomly, were replaced by new values.
The number of cells for replacement was estimated (from the average variance of the polarization degree) as follows:
\begin{equation}
 N_{\rm var}(\Delta t_{\rm i}) = \frac{\Delta t_{\rm i}}{\langle \Delta t \rangle} \frac{\sigma (P_{\rm obs})}{\langle P_{\rm obs} \rangle} N,
\end{equation}
where $\sigma (P_{\rm obs})$ is the observed standard deviation of the degree of polarization for each blazar, and
$\langle \Delta t \rangle$ is the average time difference between observations.

It was confirmed that the simulated and observational data in corresponding blazars have similar statistical properties.
Namely, the standard deviation and average of the polarization fraction are consistent with $\sigma (P_{\rm obs})$ and
$\langle P_{\rm obs} \rangle$.

\subsubsection{Individual rotations} \label{subsec:MC_simul_ind}

Using the algorithm described in Sec.~\ref{sec:rotations}, i.e. the same algorithm we used to identify rotations in real
data, we identified all rotations in the simulated data and found the number $N_{\rm rot}$ of ``successful'' data sets,
where at least one rotation with $\Delta \theta_{\rm max}$ larger or equal to $\Delta \theta_{\rm max,obs}$ was detected.
We then estimated two ratios: $P(\rm{RW})=N_{\rm rot}/10^4$ and $T_{\rm occ}=10^4 \cdot T_{\rm obs}/N_{\rm rot}$. The first
ratio determines the probability to observe an EVPA rotation due to a random walk for each one of the observed EVPA curves
for the given $T_{\rm obs}$. The second ratio determines the average time interval between random walk rotations (i.e. the
average occurrence rate for each blazar). The probabilities $P(\rm{RW})$ and $T_{\rm occ}$ are listed in Table~\ref{tab:rbpl_rotations2}.
The probabilities are larger than 10\% in all but one object, and in some cases, they approach unity. This result
indicates that the rotations we observed in some objects could be the result of a random walk process.

\subsubsection{Rotations as a population} \label{subsec:MC_simul_pop}

In this section we test the hypothesis that {\em all} the rotations observed by RoboPol in blazars of the main sample
are produced by the stochastic process. According to the analysis in Sec.~\ref{subsec:dist_swings} blazars exhibiting
rotations have different properties when compared to non-rotators. Therefore the sample of rotators must be considered
separately.

We performed the following simulation. At each iteration, an artificial EVPA curve was generated individually for each 
rotator from the main sample as explained in Sec.~\ref{subsec:MC_simul}. In each of the simuated EVPA curves we identified
the largest rotation and constructed the CDF of $\Delta \theta_{\rm max,simul}$ among the blazars. An iteration was
considered to be ``successful'' only in the case when the CDF of $\Delta \theta_{\rm max,simul}$ was lower or equal to
the CDF of $\Delta \theta_{\rm max,obs}$, i.e. the simulated set of EVPA curves had higher or equal fraction of rotations
of a given length compared to the observed set. In the cases of RBPLJ1806+6949 and RBPLJ2232+1143 where double rotations
were observed, we simulated only the largest $\Delta \theta_{\rm max}$ rotations.

\begin{figure}
 \centering
 \includegraphics[width=0.39\textwidth]{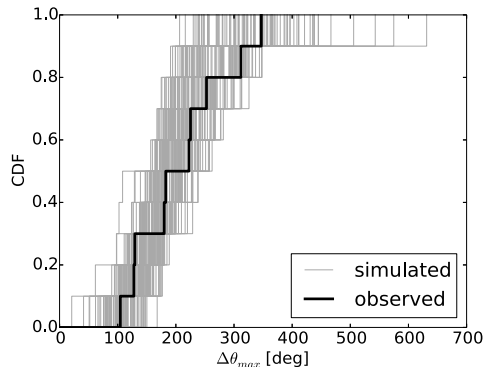}
\caption{CDFs of $\Delta \theta_{\rm max}$ in observed and a subset of 100 simulated rotations.}
 \label{fig:sim_CD_rot}
\end{figure}

The CDF of $\Delta \theta_{\rm max,obs}$ along with a subset of 100 simulated CDFs is shown in Fig.~\ref{fig:sim_CD_rot}.
It was found that only $1.5\%$ in $10^4$ trials were ``successful''. Therefore, the probability that 10 largest
rotations in blazars of the main sample observed in our monitoring campaign {\em all together} were produced by a random
walk is $\sim 1.5\%$. If we repeat this simulation for the whole set of 16 EVPA rotations this probability is reduced
to $0.5\%$.

We conclude that, although some of the rotation events that we have detected may have been caused by a random walk process
(as it is modelled in this paper), this hypothesis is not a likely explanation of the total number of detected EVPA rotations
in our data set.

\begin{figure*}
 \centering
 \includegraphics[width=0.44\textwidth]{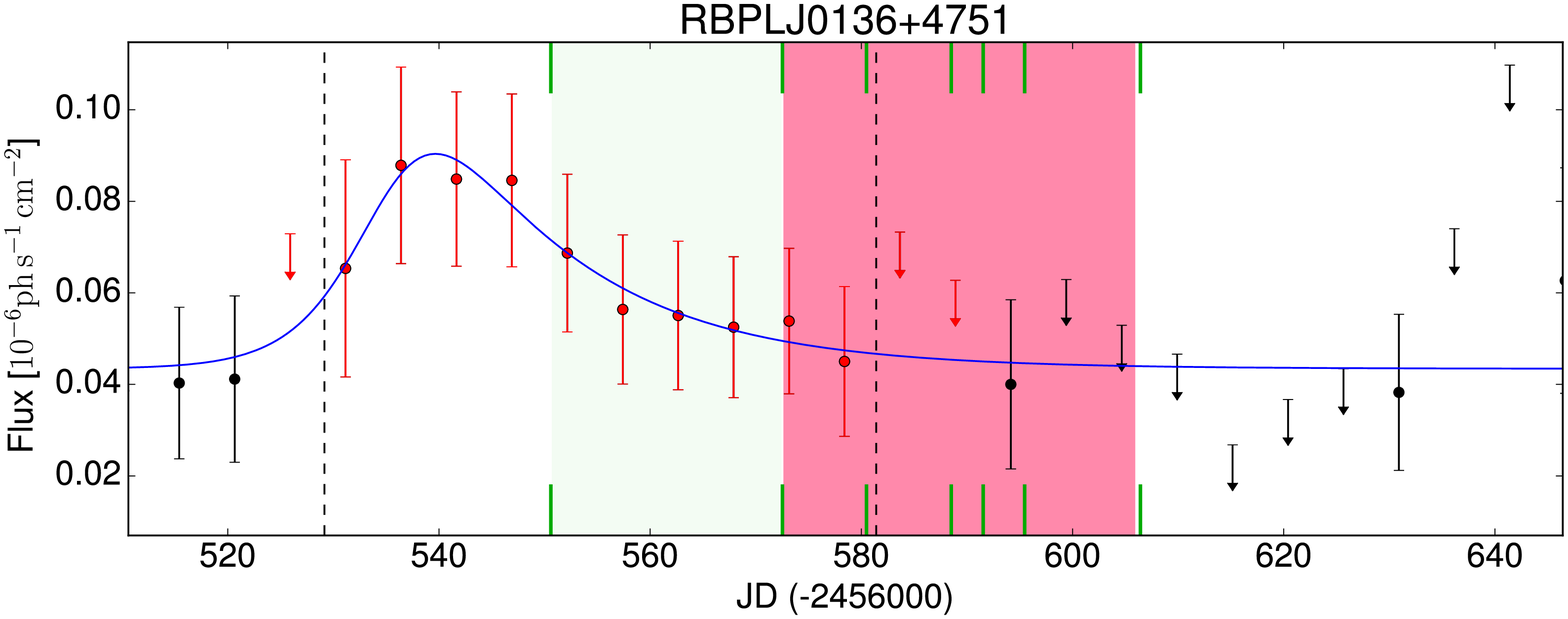}
 \includegraphics[width=0.44\textwidth]{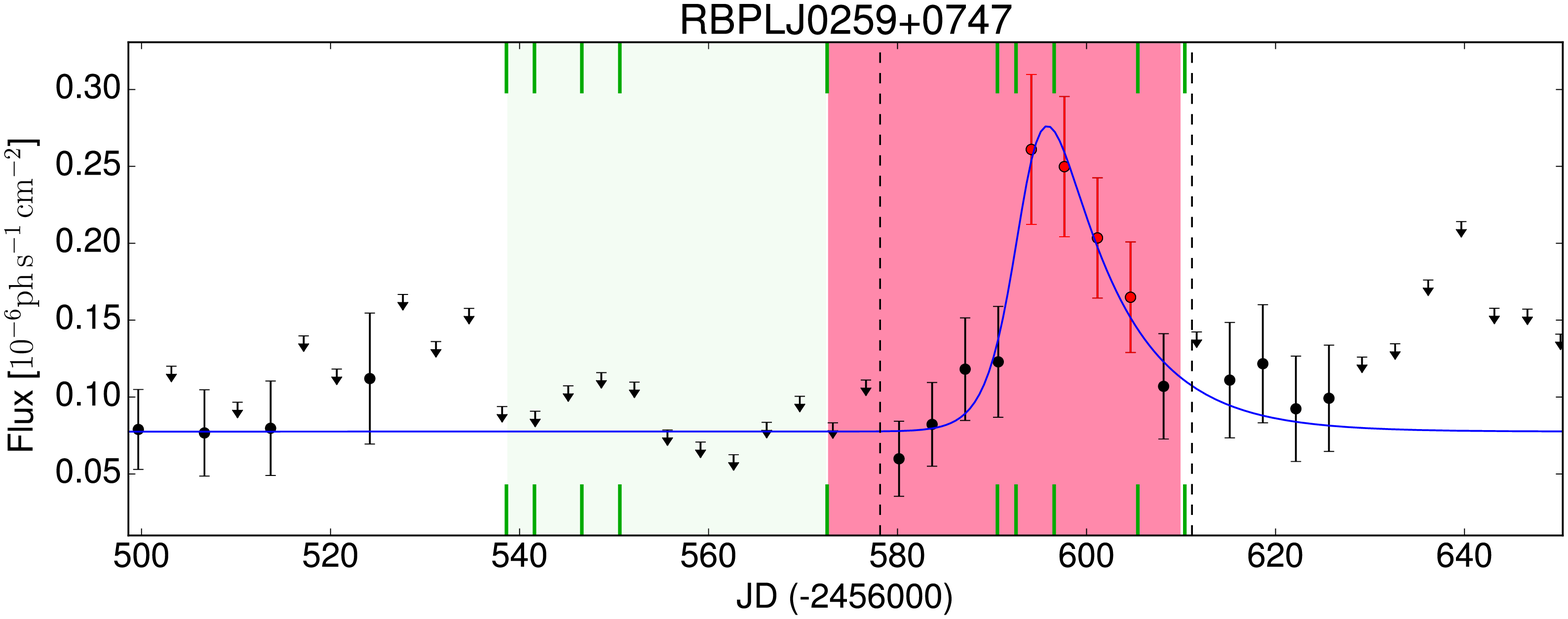}
 \includegraphics[width=0.44\textwidth]{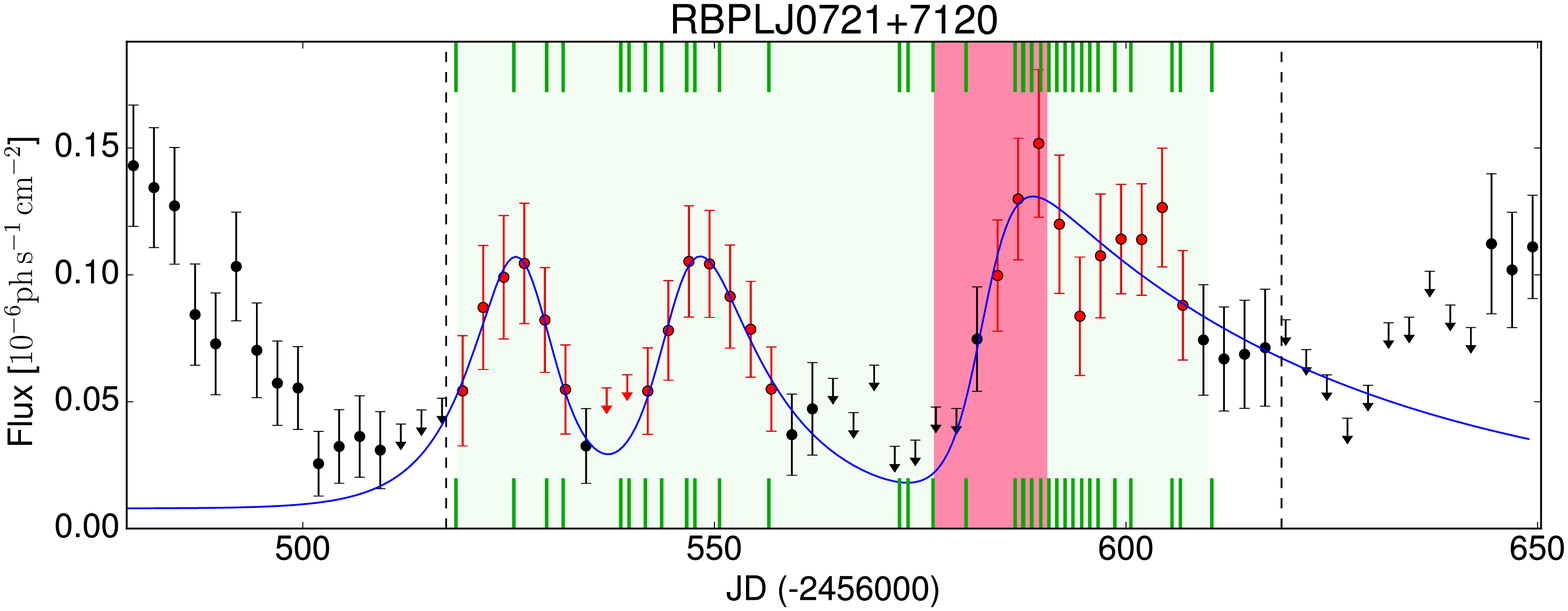}
 \includegraphics[width=0.44\textwidth]{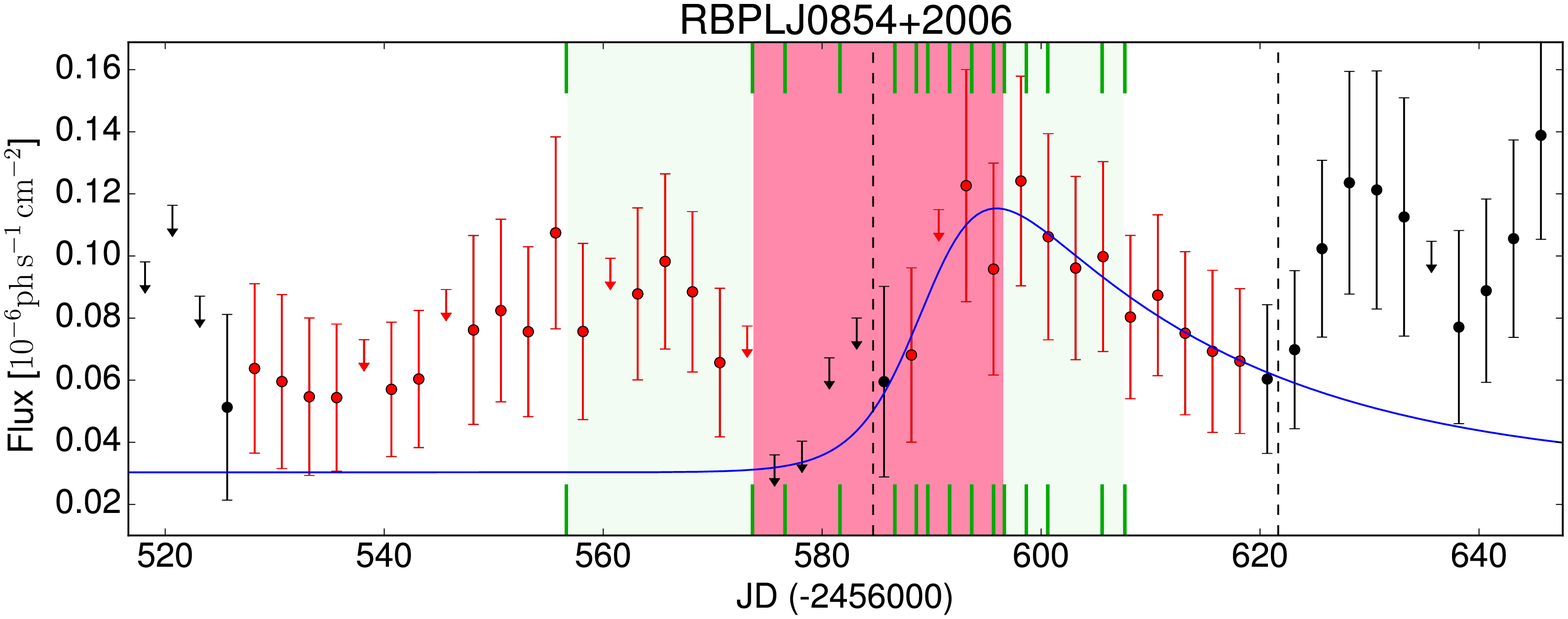}
 \includegraphics[width=0.44\textwidth]{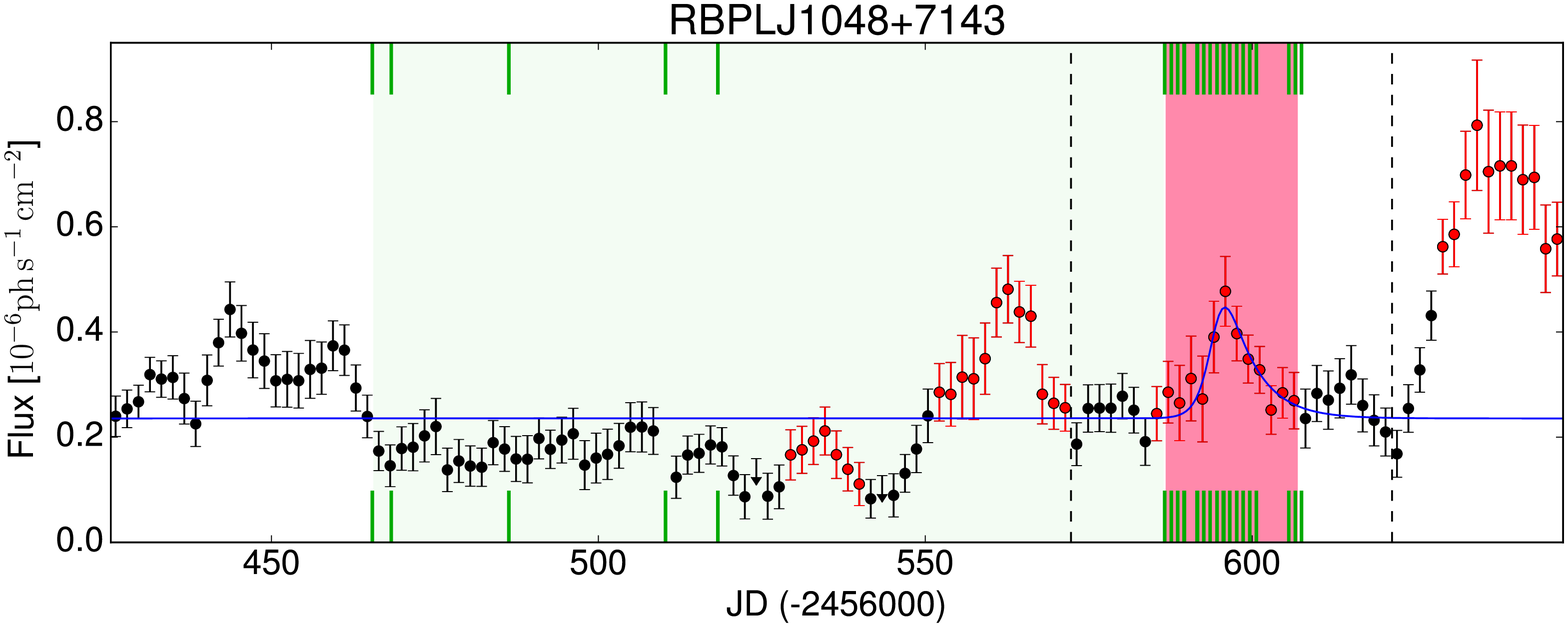}
 \includegraphics[width=0.44\textwidth]{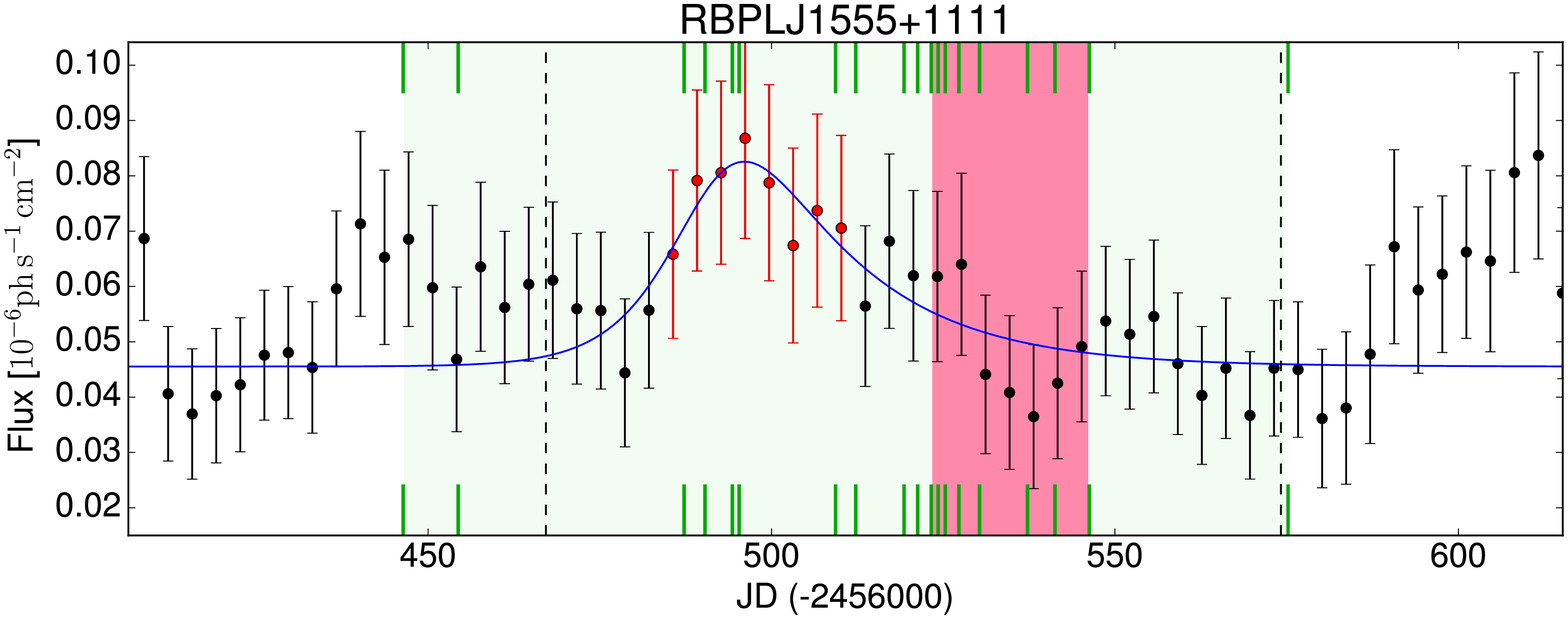}
 \includegraphics[width=0.44\textwidth]{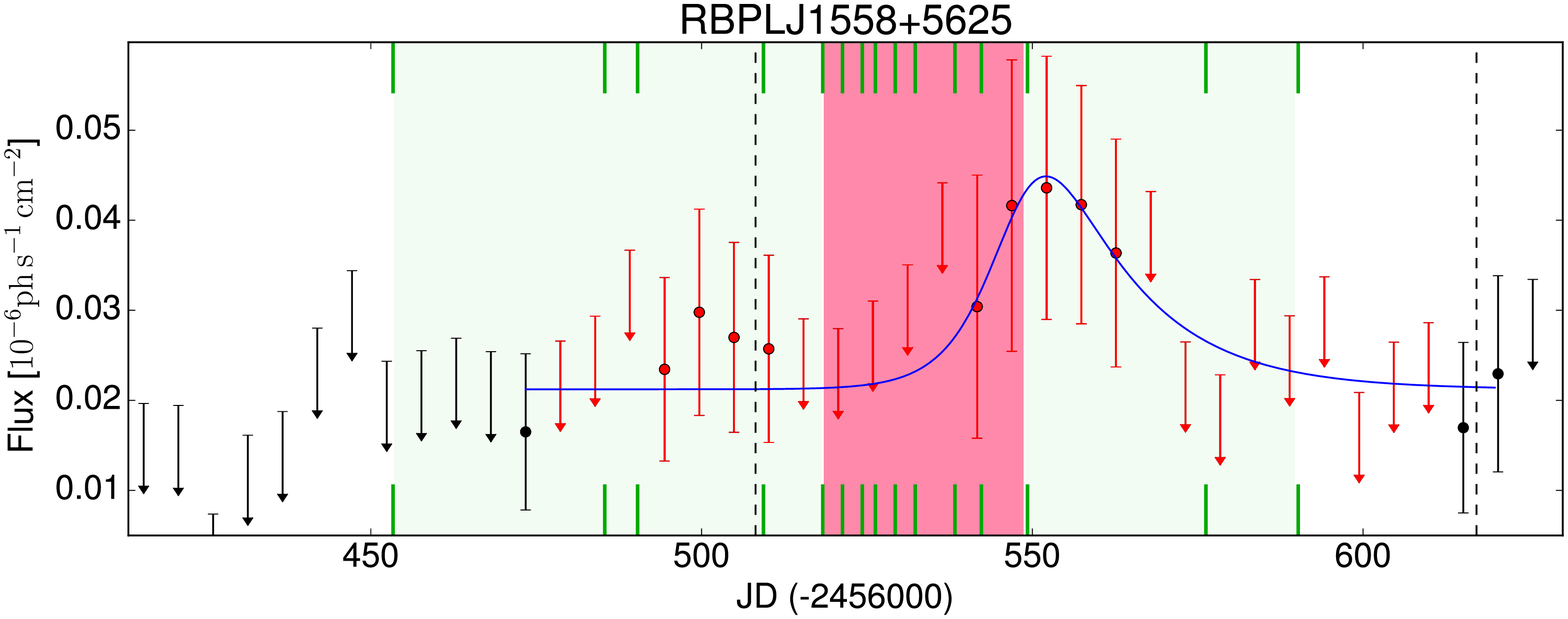}
 \includegraphics[width=0.44\textwidth]{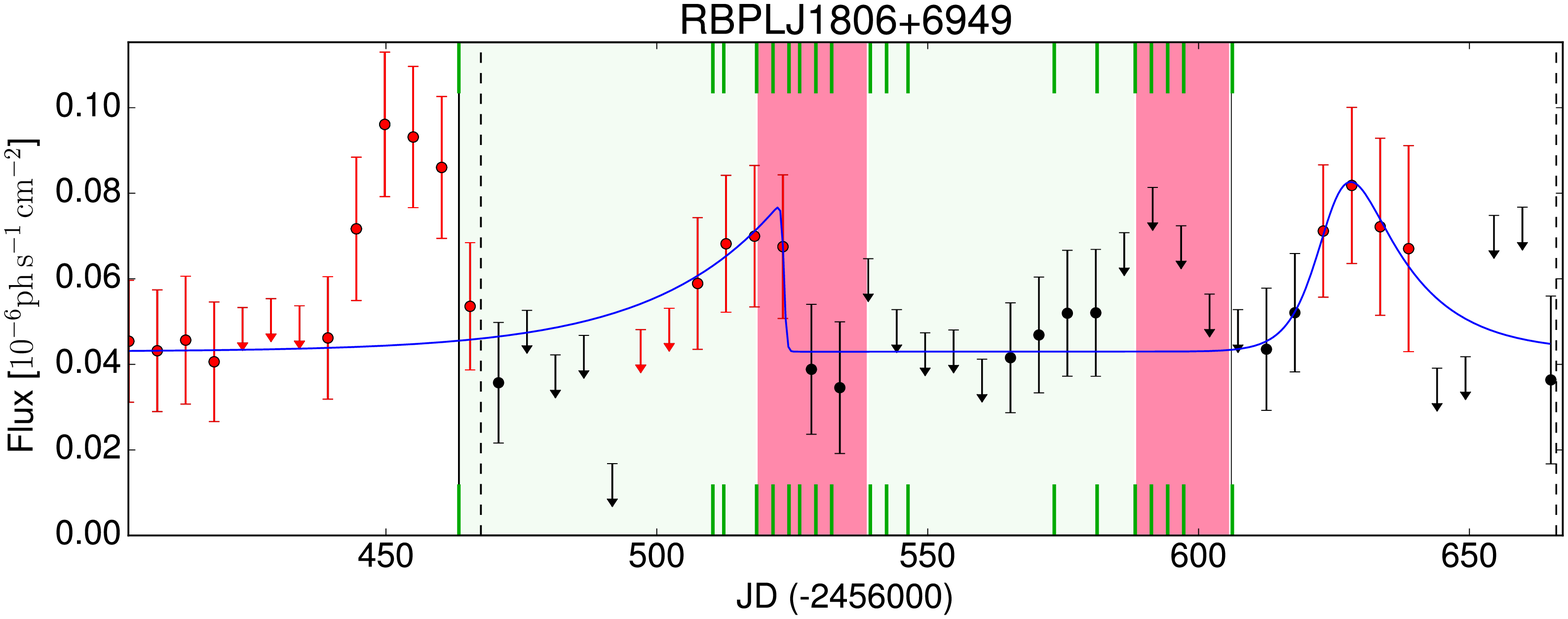}
 \includegraphics[width=0.44\textwidth]{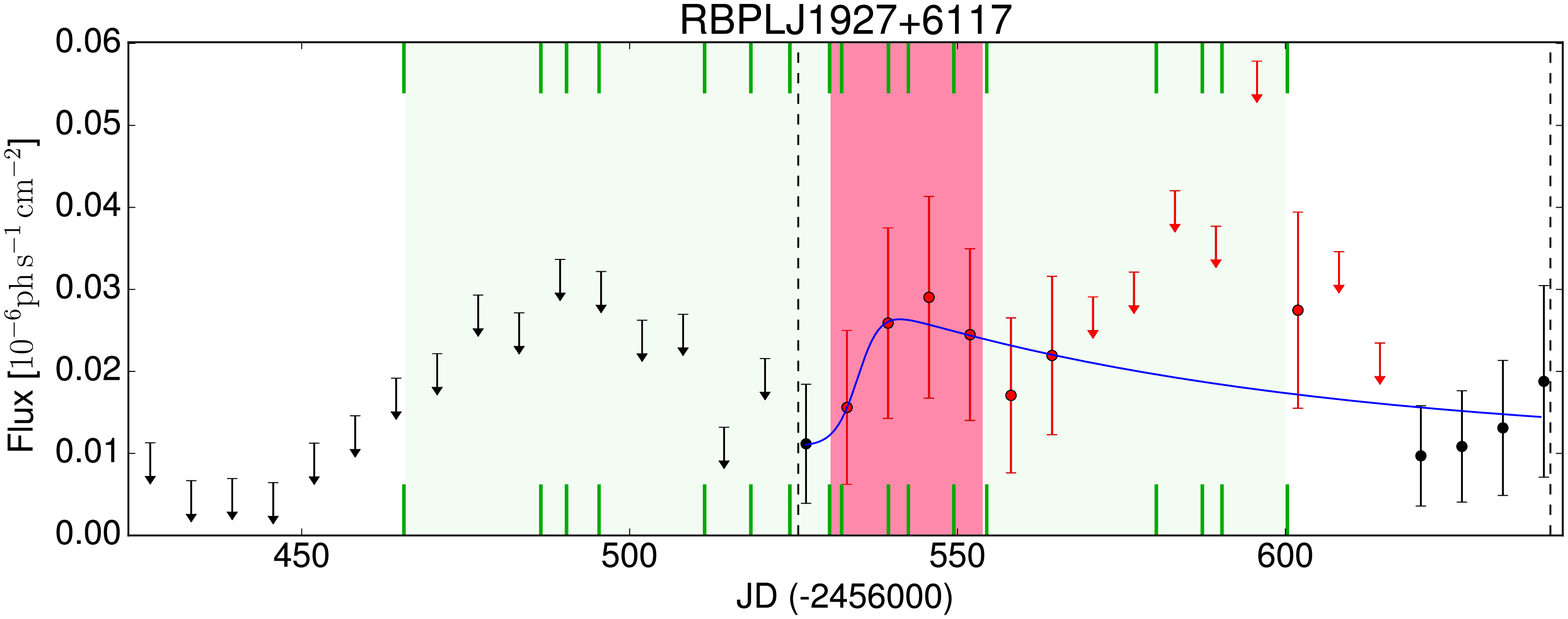}
 \includegraphics[width=0.44\textwidth]{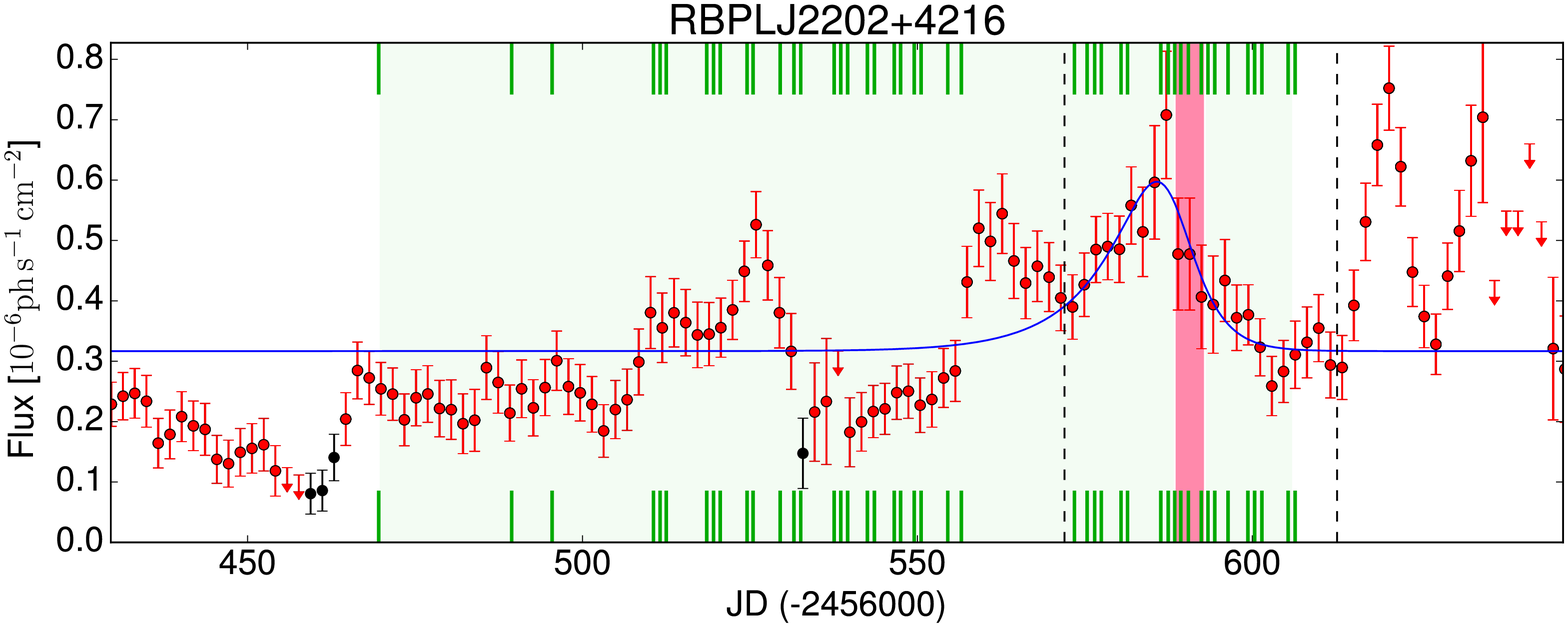}
 \includegraphics[width=0.44\textwidth]{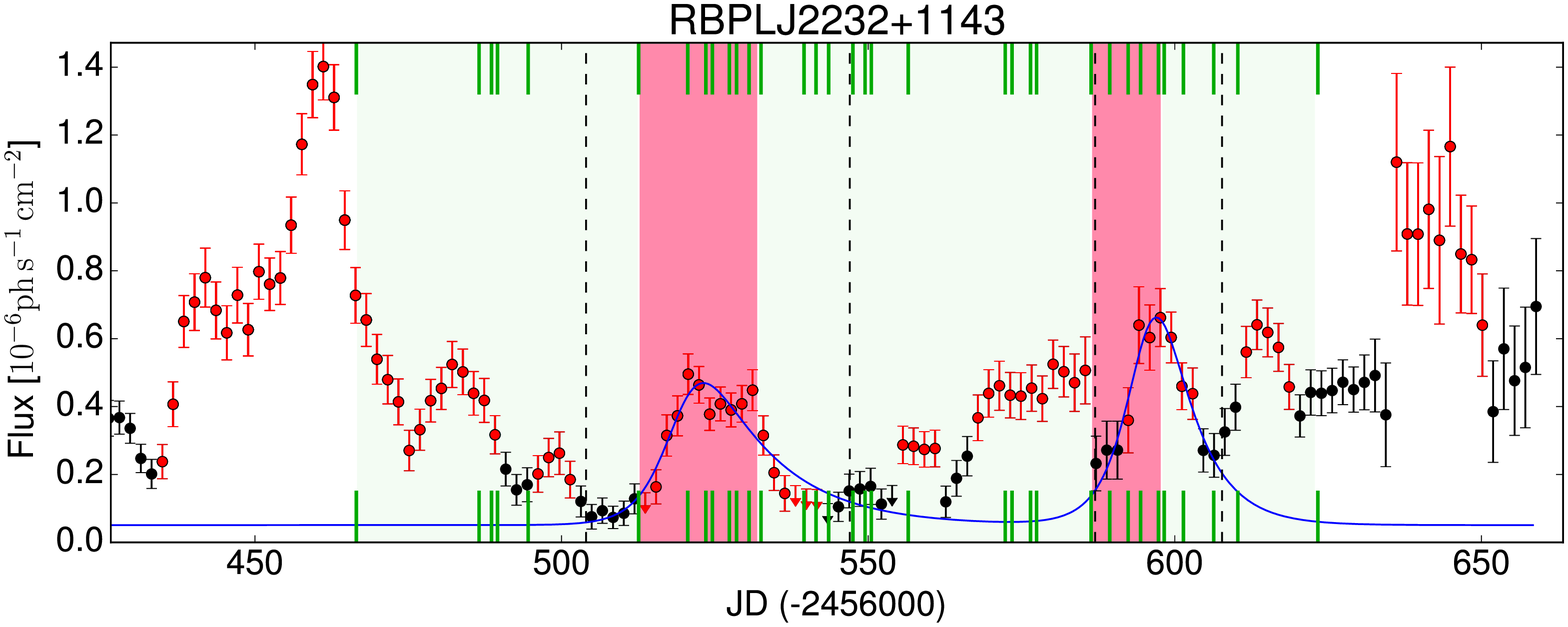}
 \includegraphics[width=0.44\textwidth]{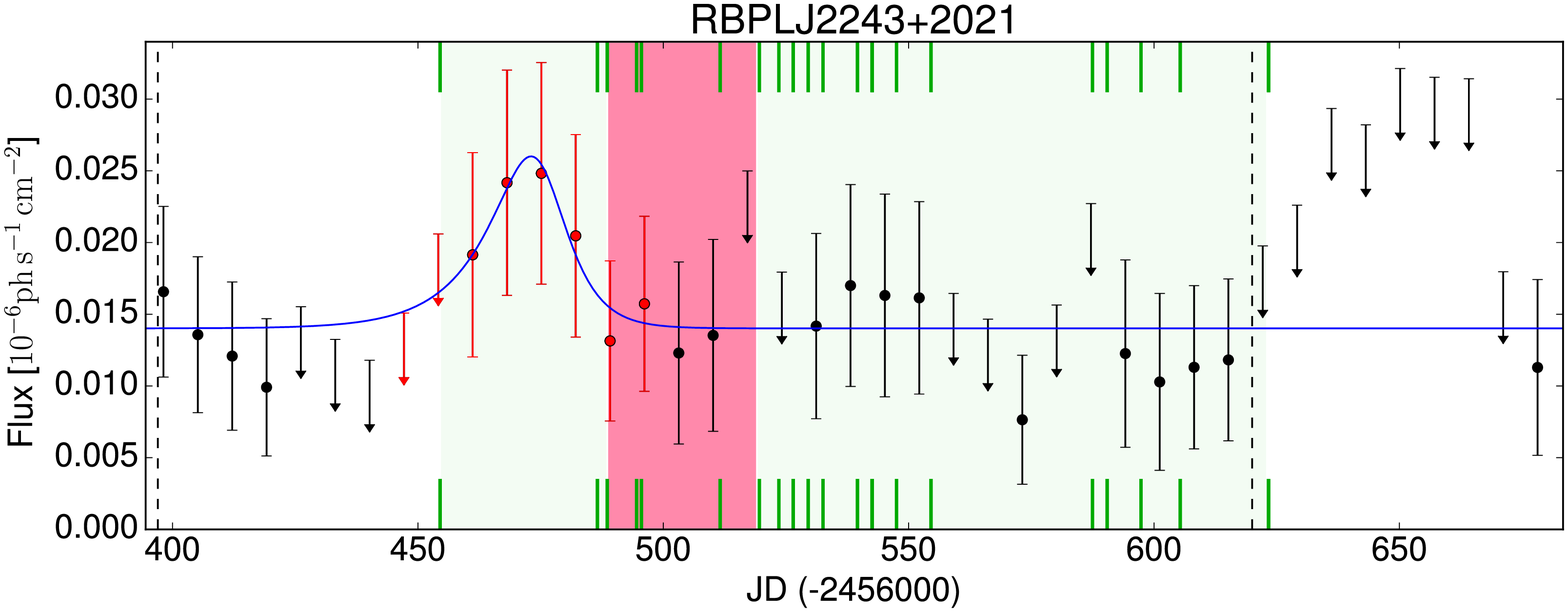}
 \includegraphics[width=0.44\textwidth]{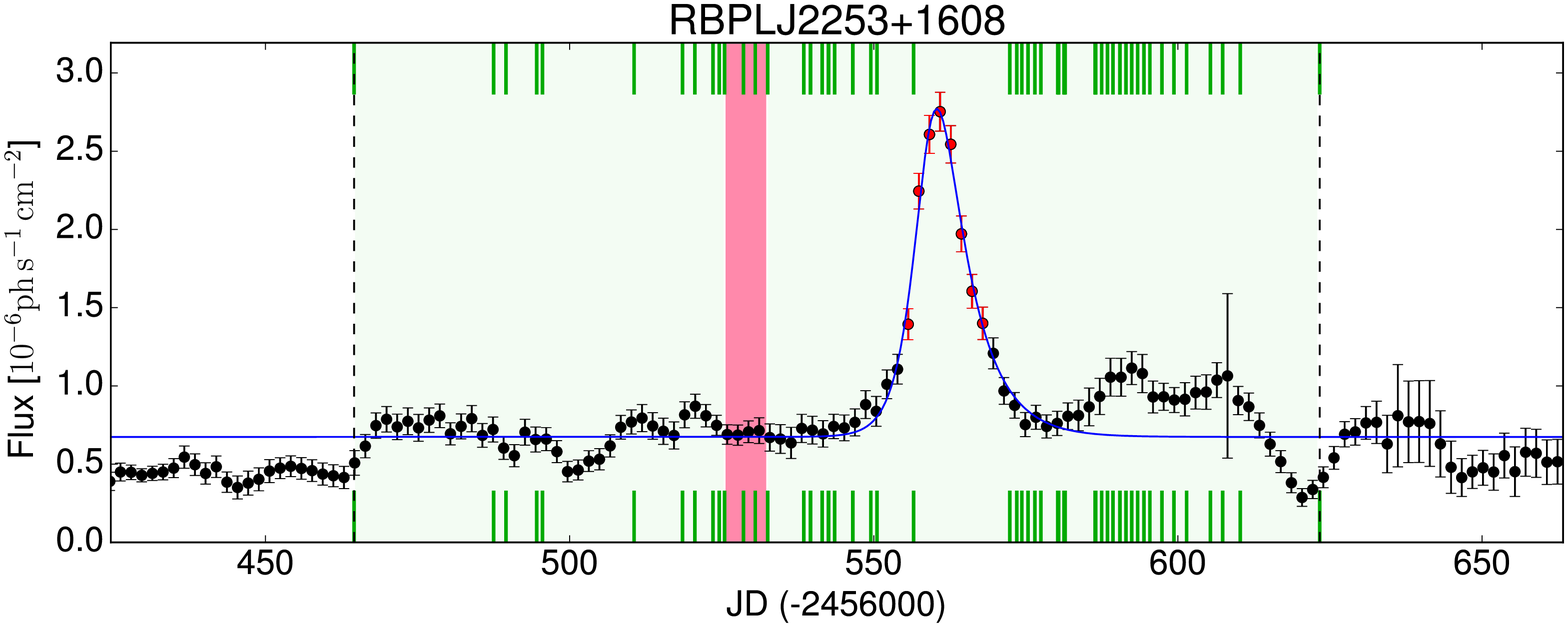}
 \includegraphics[width=0.44\textwidth]{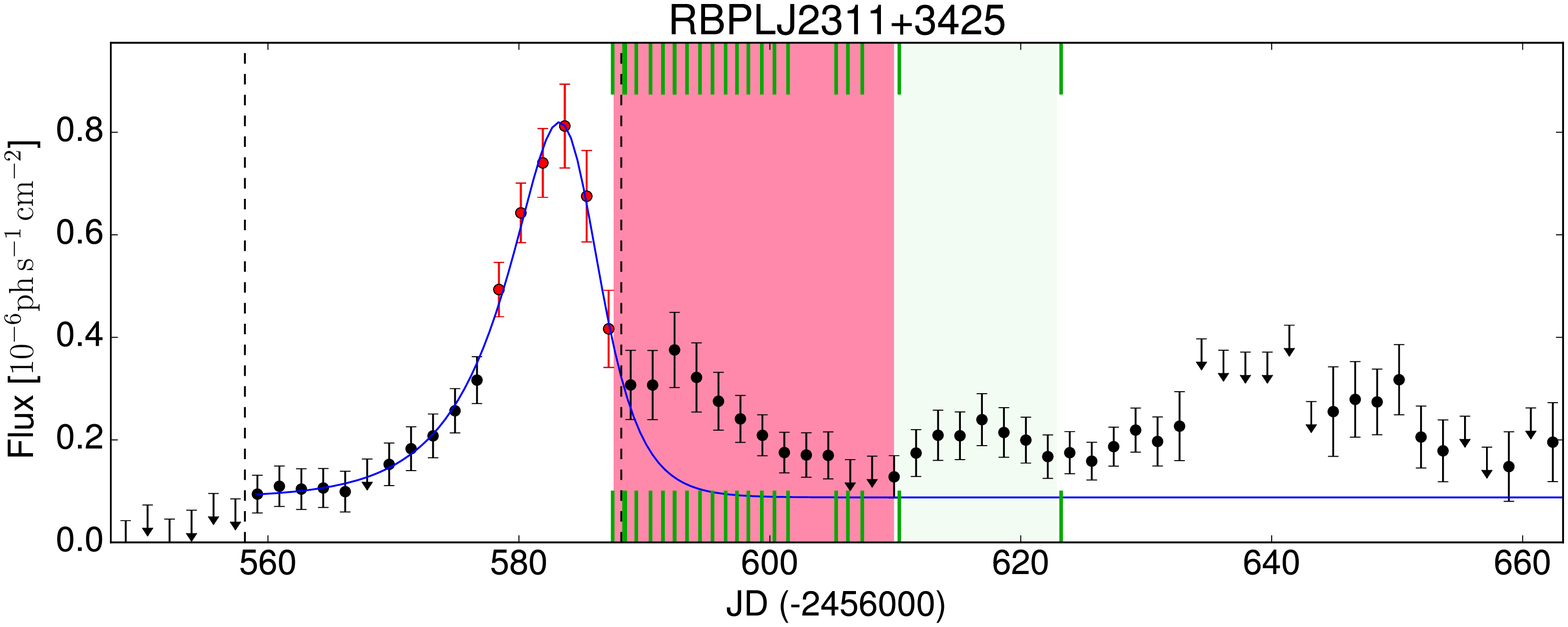}
\caption{Gamma-ray light curves of objects with detected rotations of EVPA. The RoboPol observational season is marked by
the green (light) area. The pink (dark) area shows duration of the rotation. Green ticks mark moments of our optical EVPA
measurements. All curves are centred to the mean day of the RoboPol observing season. Detected flares are marked by red
points, while the blue curve is the analytical function fit of the flares closest to observed rotations (see text for
details). Vertical dashed lines indicate intervals of the light curves used in the fitting procedure.}
 \label{fig:gamma_rotations}
\end{figure*}

\section{Optical EVPA rotations and gamma-ray activity} \label{sec:gamma}
\subsection{Average gamma-ray flux during EVPA rotations} \label{subsec:average_flux}
It has been suggested \citep[e.g.][]{Abdo2010,Marscher2010} that rotations of EVPA in optical emission of blazars are
physically related to gamma-ray flares.

In order to quantify a possible connection between EVPA rotations and gamma-ray activity, we first compared
the average gamma-ray photon fluxes for each blazar during rotation events with the rest of the RoboPol
season where no rotation was detected. Fig.~\ref{fig:gamma_rotations} shows the gamma-ray lightcurves of blazars with
detected rotations of EVPA. The green (light) area indicates the first year RoboPol observational season for each object
and the pink (dark) area indicates the period of the detected rotation. The average photon fluxes (listed in
Table~\ref{tab:gamma_flux}) were calculated using the time intervals corresponding to the rotating and non-rotating periods
as single time bins (or averaging fluxes for two/three non-rotating time bins in cases, where they are split by the rotations).
The gamma-ray photon flux during a rotation was higher than the flux during the rest
of the season at 1-$\sigma$ level only in four cases. The average difference between the photon flux during rotations and
along the rest of the season is $-0.3\pm3.4\times10^{-9} {\rm ph\, cm^{-2}\, s^{-1}}$. Thus, we do not observe any significant
systematic change of the average gamma-ray photon flux simultaneous with the EVPA rotations.

However, a comparison of the mean flux levels during the rotation and over a relatively long period may not be the best
way to search for a correlation between the gamma-ray activity and EVPA rotations. For instance in the cases of RBPLJ0721+7120
and first rotation of RBPLJ2232+1143, rotations are clearly coincident with prominent flares, although the average
gamma-ray photon fluxes are indistinguishable, since the season comprises a number of flaring events with similar amplitude.
Moreover, rotations of EVPA can either precede or follow gamma-ray flares according to various theoretical scenarios. It
is therefore important to search for a possible correlation between EVPA rotations and gamma-ray flares.

\begin{table}
 \centering
  \caption{Gamma-ray photon flux level during rotations and throughout the rest of the RoboPol season.}
  \label{tab:gamma_flux}
  \begin{tabular}{@{}ccc@{}}
  \hline
             &   \multicolumn{2}{c}{Photon flux ($E > 100$ MeV)}      \\
  Blazar ID  &   \multicolumn{2}{c}{($10^{-7} {\rm ph \,cm^{-2} s^{-1}}$)}  \\
             &  rotation   &   no rotation \\
 \hline
RBPLJ0136+4751 & $0.40 \pm 0.14$ & $0.59 \pm 0.16$ \\
RBPLJ0259+0747 & $1.27 \pm 0.21$ & $<0.71$ \\
RBPLJ0721+7120 & $0.95 \pm 0.18$ & $0.84 \pm 0.11$ \\
RBPLJ0854+2006 & $0.33 \pm 0.16$ & $0.91 \pm 0.18$ \\
RBPLJ1048+7143 & $3.39 \pm 0.32$ & $2.12 \pm 0.11$ \\
RBPLJ1555+1111 & $0.51 \pm 0.11$ & $0.54 \pm 0.05$ \\
RBPLJ1558+5625 & $<0.34$         & $0.21 \pm 0.05$\\
RBPLJ1806+6949 & $0.35 \pm 0.15$ & $0.40 \pm 0.07$ \\
RBPLJ1806+6949 & $<0.83 $        & $0.40 \pm 0.07$ \\
RBPLJ1927+6117 & $0.29 \pm 0.13$ & $0.09 \pm 0.05$\\
RBPLJ2202+4216 & $4.67 \pm 0.93$ & $3.29 \pm 0.21$ \\
RBPLJ2232+1143 & $3.82 \pm 0.32$ & $3.34 \pm 0.25$ \\
RBPLJ2232+1143 & $4.55 \pm 0.70$ & $3.34 \pm 0.25$ \\
RBPLJ2243+2021 & $0.11 \pm 0.06$ & $0.17 \pm 0.04$\\
RBPLJ2253+1608 & $6.98 \pm 0.68$ & $8.82 \pm 0.19$ \\
RBPLJ2311+3425 & $2.09 \pm 0.27$ & $2.13 \pm 0.35$ \\
\hline
\end{tabular}
\end{table}

\subsection{Time lags between flares and EVPA rotations} \label{subsec:flare_lag}

In order to investigate this relation, we first identified all flares that happened in the gamma-ray light curves during
the RoboPol observing season.

We adopted a formal definition of a gamma-ray flare similar to the one proposed by \cite{Nalewajko2013}: ``a flare is a
contiguous period of time, associated with a given photon flux peak, during which the photon flux exceeds half of the peak
value, and this lower limit is attained exactly twice -- at the start and at the end of the flare''. However the definition
was slightly changed because \cite{Nalewajko2013} analysed a sample of the brightest flares ever detected by {\em Fermi}
LAT, while we are interested in even smaller amplitude events. We found that a peak photon flux excess factor equal to
$2/3$, instead of the original $1/2$ proposed by \cite{Nalewajko2013}, gives a better agreement with a visual identification
of flares in the photon flux curves. Intervals of the photon flux curves identified as flares are marked by red (light)
points in Fig.~\ref{fig:gamma_rotations}.

We searched for the closest gamma-ray flare to the rotation event of each rotator, and we fitted it using a profile with
an exponential rise and decay. This kind of profile is commonly used for fitting an individual blazar flare pulse in
optical, gamma and radio bands \citep[e.g.][]{Abdo2010b,Chatterjee2012}:
\begin{equation}\label{fit_func}
 F(t) = F_{\rm c} + F_{\rm p} \left( e^{\frac{t_{\rm p}-t}{T^{\rm r}}} + e^{\frac{t-t_{\rm p}}{T^{\rm d}}} \right)^{-1},
\end{equation}
where $F_{\rm c}$ represents an assumed constant level underlying the flare, $F_{\rm p}$ measures the amplitude of the
flare, $t_{\rm p}$ describes the time of the peak (it corresponds to the actual maximum only for symmetric flares), $T^{\rm r}$ and
$T^{\rm d}$ measure the rise and decay time, respectively. All the parameters were set to be free, while initial values used
in the fitting procedure were estimated from the photon flux curves. Upper limits of the light curve were not used
in the fitting procedure. In the case of RBPLJ0721+7120 three flares that occurred during the observing season were fitted
simultaneously because a single flare fit resulted in an unrealistic $F_{\rm c}$ value. In the cases of double rotations
in RBPLJ1806+6949 and RBPLJ2232+1143 the flares closest to the rotations were also fitted together to provide a consistent
$F_{\rm c}$ value. In addition, in three cases, the closest flares happened just outside the RoboPol season interval. The best
fitting curves are shown in Fig.~\ref{fig:gamma_rotations}.

\begin{figure}
 \centering
 \includegraphics[width=0.37\textwidth]{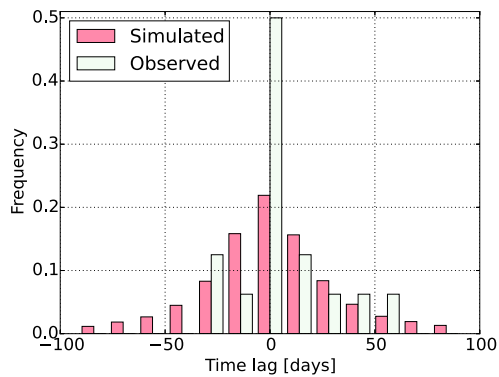}
\caption{Distributions of observed $\tau_{\rm obs}$ and simulated $\tau_{\rm simul}$ time lags between middle points of
rotations and $t_{\rm p}$ of gamma-ray flares.}
 \label{fig:tl_dist}
\end{figure}

We estimated time lags, $\tau_{\rm obs}$, between rotations and the closest gamma-ray flares as
$\tau_{\rm obs} = \overline{T^{\rm rot}} - t_{\rm p}$, where $\overline{T^{\rm rot}}$ is the middle point of each EVPA
rotation, defined as $t_{\rm  rot,start} + \frac{T_{\rm rot}}{2}$ (see Table \ref{tab:rbpl_rotations3}). The time lags
have a distribution, shown by green (light) bars in Fig.\ref{fig:tl_dist}, with mean and standard deviation equal to 5.1
and 21.8 d, respectively. The distribution is indistinguishable from the normal distribution $N(0,21.8)$ following
the K-S test ($p\text{-value} = 0.39$). Thereby {\em we do not find any preference for positive or negative $\tau_{\rm obs}$}.
A distribution of observed time lags is expected to be close to a normal distribution with the mean at zero if rotations
of the EVPA are not connected to gamma-ray flares. Because, in this case, the overall distribution is produced by a set
of random values each having distributions of different widths and symmetric with respect to zero. For this reason the
time lags distribution does not, on its own, support a physical connection between gamma-ray flares and rotations. However
theoretical models allow for either positive or negative lags, depending on conditions and emission region properties,
when a physical connection between rotations and gamma-ray flares does exist \citep{Zhang2014}. Therefore, a physical
connection cannot be excluded based on the distribution of $\tau_{\rm obs}$.

\begin{table}
\centering
\caption{Gamma-ray flares fitting results. (1) - blazar identifier; (2) - time difference, $\tau_{\rm obs}$, between 
$t_{\rm p}$ of the closest gamma-ray flare and
middle point of the rotation (positive means leading flare); (3) - gamma-ray flare amplitude measured relative to the
average photon flux of the blazar from 2FGL.}
\label{tab:rbpl_rotations3}
  \begin{tabular}{lcc}
  \hline
   Blazar ID   & $\tau_{\rm obs}$ & $\gamma$-flare   \\
               & (d)              & rel. ampl.       \\
 \hline
RBPLJ0136+4751 &    $53.8$    &   $0.6  \pm 0.08$ \\
RBPLJ0259+0747 &   $-$2.4     &   $15.1 \pm 2.9$ \\
RBPLJ0721+7120 &      0.8     &   $1.0  \pm 0.5$ \\
RBPLJ0854+2006 &   $-$5.3     &   $2.5  \pm 1.1$ \\
RBPLJ1048+7143 &      2.5     &   $7.3  \pm 3.6$ \\
RBPLJ1555+1111 &     42.9     &   $1.1  \pm 0.2$ \\
RBPLJ1558+5625 &   $-$14.8    &   $1.9  \pm 0.9$ \\
RBPLJ1806+6949 &      5.4     &   $0.7  \pm 0.6$ \\
RBPLJ1806+6949 &    $-$27.8   &   $1.3  \pm 0.4$ \\
RBPLJ1927+6117 &      7.5     &   $0.6  \pm 0.2$ \\
RBPLJ2202+4216 &     3.1      &   $3.1  \pm 0.6$ \\
RBPLJ2232+1143 &      2.6     &   $7.3  \pm 1.5$ \\
RBPLJ2232+1143 &  $-$3.7      &   $12.1 \pm 1.5$ \\
RBPLJ2243+2021 &    29.0      &   $0.7  \pm 0.4$ \\
RBPLJ2253+1608 &   $-$30.2    &   $1.7  \pm 0.2$ \\
RBPLJ2311+3425 &    18.8      &   $16.6 \pm 1.3$ \\
\hline
\end{tabular}
\end{table}

\subsection{Relation of gamma-ray flare amplitudes and time delays} \label{subsec:gamma_rot_propert}

\begin{figure}
 \centering
 \includegraphics[width=0.48\textwidth]{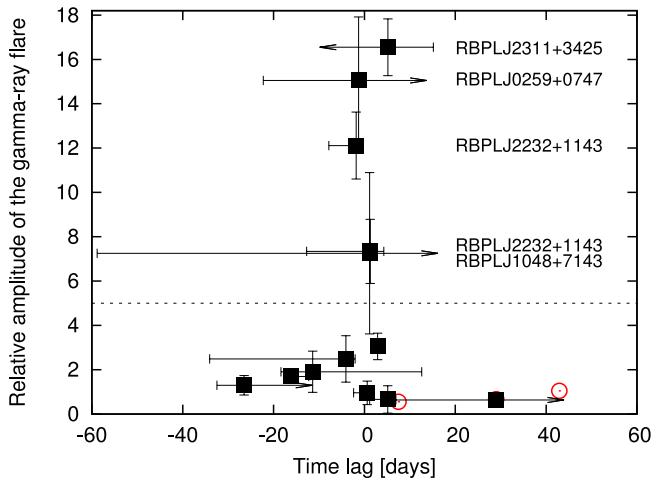}
\caption{Time lags, $\tau_{\rm obs}$, versus normalized gamma-ray flare amplitude, $F_{\rm p}$. Redshift corrected and
non-corrected $\tau_{\rm obs}$ values are plotted with filled squares and open circles, respectively.}
 \label{fig:tl_vs_av_pol}
\end{figure}

We normalized the amplitude, $F_{\rm p}$, of the gamma-ray flare closest to the EVPA rotation event by the average photon flux
of each blazar \citep[as listed in 2FGL;][]{Nolan2012}. Corresponding values are listed in Table~\ref{tab:rbpl_rotations3}
and plotted as a function of $\tau_{\rm obs}$ in Fig.~\ref{fig:tl_vs_av_pol}. The filled
black squares show redshift-corrected time lags, i.e. $\tau_{\rm corr} = \tau_{\rm obs}/(1+z)$, while open circles show
$\tau_{\rm obs}$ for blazars with unknown $z$. The ``errors'' on the time lags are defined as fitting errors of $t_{\rm p}$
plus the time difference between the first/last point of a rotation event, and the previous/next closest point of the EVPA
curve. Due to the lack of data in some cases these ``uncertainties'' are undefined, while in others, due to sparse sampling,
they are almost certainly overestimated.

A noticeable feature is that $\tau_{\rm corr}$ is in the range $(-6,+6)$ d for the most prominent gamma-ray flares. Basically,
all five brightest flares have happened almost simultaneously with EVPA rotation events. The brightest flare which
has the largest deviation from the zero-delay is the flare of RBPLJ2311+3425 where the start point of the rotation is
undefined and therefore the time-delay has a large uncertainty.

There are three more flares with similarly small time lags, and small relative amplitudes. Thus a small separation between
a flare and a rotation is not a sufficient condition for extraordinary brightness of the high-energy flare.

Separating the flares into two subsamples of high and low amplitude events (dashed line in Fig.~\ref{fig:tl_vs_av_pol})
we examined the significance of the difference in time delays between them. The mean of the absolute $\tau_{\rm obs}$ values
for the high and low amplitude subsamples is 5.2 and 20.1 d, respectively. According to the Student's {\em t}-test 
\citep[e.g.][]{Wall2012}, the difference between the two mean values is somewhat significant ($p\text{-value}=0.025$).

\subsection{Accidental proximity of rotations and gamma-ray flares} \label{subsec:lag_prob}
\subsubsection{Individual blazars}

In order to estimate the probability of the accidental observed proximity in time of rotations and gamma-ray
flares, we performed MC simulations using the observed gamma-ray photon flux curves. This allows us to
account for the real variability of blazars in the gamma-ray band. For each rotator we processed a long-term set of
{\em Fermi} LAT data ($54683 \le {\rm MJD} \le 57065$) with time bins equal to the ones used in Sec.~\ref{subsec:flare_lag}.
Then we identified and fitted all gamma-ray flares following the procedure described previously, using
the same photon flux excess factor of 1.5. The number of flares identified in the photon flux curves of rotators is in
the range of 12 - 76. After that we randomly assigned the middle point of a simulated rotation to a time on the photon
flux curve and measured the time lag between the rotation and the closest gamma-ray flare, $\tau_{\rm simul}$. Repeating
this simulation $10^4$ times for each blazar, we determined the distributions of time delays $\tau_{\rm simul}$. Using
these distributions, we estimated the probability of $\tau_{\rm obs}$ to be produced by chance $P(\tau_{\rm obs})$, by
calculating the fraction of simulations where $\tau_{\rm simul} \le \tau_{\rm obs}$. The probabilities range between
3 and 78 per cent (see Table~\ref{tab:model_delays}). Pink (dark) boxes in Fig.~\ref{fig:tl_dist} indicate the
distribution of $\tau_{\rm simul}$, using the results from the simulation for all blazars. According to the K-S test the
null hypothesis that $\tau_{\rm simul}$ and $\tau_{\rm obs}$ are drawn from the same distribution can not be rejected
($p\text{-value} = 0.38$). Therefore, it is possible that the $\tau_{\rm obs}$ values we observed, may be accidental for
each of the blazars in the sample.

In Sect.~\ref{subsec:MC_simul} we determined the probability of the EVPA rotations to be observed in our observing
window assuming that they are produced by a stochastic process. The simulations described above give us the probability
of an accidental simultaneity between these rotations and gamma-flares. Therefore the probability of superposition of both
independent events: (a) random rotation {\em and} (b) random proximity to a gamma-ray flare, can be estimated as a product
of the respective probabilities. These combined probabilities are less than 5 per cent for five events (see column 3 of
Table~\ref{tab:model_delays}). This result indicates that, at least for some rotations, the random walk model
{\em and} the absence of any physical connection between the EVPA variability and high-energy activity is an unfavourable
interpretation.

\subsubsection{Rotators as a population}

In order to assess the probability that the {\em entire set} of the time lags appeared in the main sample rotators in a
random way, we run the following simulation. Repeating the procedure described in Sec.~\ref{subsec:flare_lag} we identified and
fitted all flares in the gamma-ray photon flux curve ($54683 \le {\rm MJD} \le 57065$) of each blazar from the main sample
with a detected rotation. Then placing a simulated rotation at a random position on each of the gamma-ray curves, we
defined the shortest time lag between the central point of the rotation and $t_{\rm p}$ of the nearest flare. After this
the CDF of absolute values of the simulated time lags was constructed for the set of 14 events.

\begin{table}
\centering
\caption{Modelling results for the connection between EVPA rotations detected by RoboPol in 2013 and gamma-ray flares.
(1) - blazar identifier; (2) probability of an accidental time lag; (3) - combined probability of a rotation being
produced by the random walk and located as close to the corresponding gamma-ray flare as it was observed.}
\label{tab:model_delays}
  \begin{tabular}{lcc}
  \hline
   Blazar ID   &  $P(\tau_{\rm obs})$ & $P(\rm{RW}+\tau_{\rm obs})$ \\
               &                  &                    \\
 \hline
RBPLJ0136+4751 &  0.75           &  0.08             \\
RBPLJ0259+0747 &  0.03           &  0.02             \\
RBPLJ0721+7120 &  0.04           &  0.01             \\
RBPLJ0854+2006 &  0.23           &  0.08              \\
RBPLJ1048+7143 &  0.14           &  0.11              \\
RBPLJ1555+1111 &  0.72           &  0.72              \\
RBPLJ1558+5625 &  0.20           &  0.10              \\
RBPLJ1806+6949 &  0.10           &  0.02              \\
RBPLJ1806+6949 &  0.49           &  0.27              \\
RBPLJ1927+6117 &  0.08           &  0.08              \\
RBPLJ2202+4216 &  0.21           &  0.04              \\
RBPLJ2232+1143 &  0.14           &  0.01              \\
RBPLJ2232+1143 &  0.19           &  0.17              \\
RBPLJ2243+2021 &  0.48           &  0.44              \\
RBPLJ2253+1608 &  0.78           &  0.67              \\
RBPLJ2311+3425 &  0.56           &  0.41              \\
\hline
\end{tabular}
\end{table}

\begin{figure}
 \centering
 \includegraphics[width=0.42\textwidth]{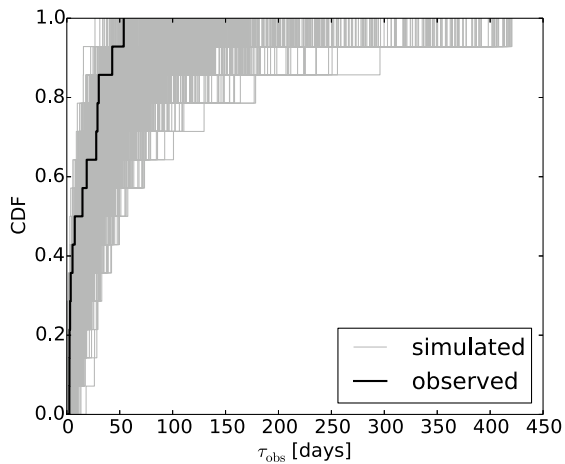}
\caption{CDFs of the time lags between the EVPA rotations middle points and $t_{\rm p}$ of
the closest gamma-ray flares for the main sample rotators. Black line -- observed time lags, thin grey lines - $10^4$
simulated values for the whole sample of rotations (see text for details).}
 \label{fig:CDF_delays}
\end{figure}

Repeating the routine $10^6$ times we found that only one out of every 5000 simulations produces a CDF which is in its
entirety located closer to zero or coincides with the CDF of observed time lags (see Fig.\ref{fig:CDF_delays}). Thereby
we estimate the probability that all 14 delays together were produced by chance as $2 \times 10^{-4}$. When we repeat
this procedure for all 16 rotations together including two non-main sample events, the estimated probability decreases to
$5 \times 10^{-5}$. Therefore, it is very unlikely that none of the observed EVPA rotations is related physically to the
flaring activity in gamma-rays.

\section{Conclusions} \label{sec:conclusion}

During the first season of operation of the RoboPol project, we detected 16 rotations of the polarization plane in optical
emission of blazars. These detections double the existing list of such events. All EVPA rotations are observed
in blazars which are detected by {\em Fermi}, in agreement with previous experiments, which have detected similar events in
the same class of objects. Our strategy of monitoring both gamma-ray--loud and quiet samples, suggests that the lack of
EVPA rotations detection by RoboPol in gamma-ray--quiet objects cannot be due to a difference in the sampling pattern.
Combining our results with those reported in the literature we found that rotations can be detected in both TeV and non-TeV
emitters. Our results also indicate that all subclasses of blazars show rotations of the EVPA (regardless of the position
of the synchrotron peak maximum or the BL Lac/FSRQ dichotomy). We expect that the results after the 3-yr planned RoboPol
monitoring campaign will allow an accurate determination of the rotations rate in the various blazar subclasses.

Analysis of the first-year data shows that blazars with detected rotations have significantly faster and longer EVPA
swings when compared to non-rotators. This suggests that rotations of EVPA may be specific for a particular activity
state or for a subclass of blazars with peculiar properties.

The fact that EVPA rotations have been detected only in gamma-ray--loud objects already suggests a physical relation
between gamma-ray and optical polarization variability in blazars. Nevertheless, we used extensive MC simulations to
investigate whether the EVPA rotations we observed can be produced by a random walk process of the polarization vector.
We found that a random walk process can result in EVPA rotations with $\Delta \theta_{\rm max,simul}$ as large as
$\Delta \theta_{\rm max,obs}$ for the given $\Delta t$-median and $T_{\rm obs}$ of the individual RoboPol data sets. However, we
also found that it is unlikely (probability is $\le 1.5 \times 10^{-2}$), that {\em all} the rotations that we observed
in the first RoboPol season are due to a random walk process.

The average gamma-ray photon fluxes do not show any significant systematic increase during the rotation events. We also
found that, the time lags between rotations of the EVPA and nearest gamma-ray flares follow a Gaussian distribution with
a mean $\sim$ zero.

We performed a second set of MC simulations in order to assess the randomness of the observed time delays.
Our results suggest that, on an individual basis, the time lags we observe do not necessarily suggest a physical link
between EVPA rotations and gamma-ray flares. On the other hand, when we consider the rotators as a population, it is
highly unlikely ($p = 2\times 10^{-4}$) that the proximity of EVPA rotations to gamma-ray flares is accidental in {\em all}
cases. Therefore at least some EVPA rotations must be physically connected to the high-energy activity. 

Our data suggest that, the highest amplitude gamma-ray flares may be physically connected with EVPA rotations, based on
the fact that they are associated with smaller-than-average time lags. Perhaps there are two different types of gamma-ray
flares, produced by different physical mechanisms. One of them may result in higher (than average) amplitude flares
{\em and} EVPA rotation events. The other one may produce the rest of the smaller amplitude flares, which are not
related with the remaining rotations, probably produced by a random walk process.

For the first time we studied a set of EVPA rotations discovered in a large, well-defined, regularly monitored sample
of blazars. The diversity of results found in individual and population analysis shows the importance of these kinds of
studies. The RoboPol monitoring of blazars will continue for at least two more years. The question about the mechanisms
responsible for the EVPA rotations in blazars and their possible connection to the high-energy activity will be explored
in more detail after accumulation of a larger data set by RoboPol.

The statistical analysis of our data set required us to make subjective choices regarding the details of our event
definitions and the test statistics we used. These, for example, include the definition of a rotation, the definition of
a ``gamma-ray activity'' (increase in gamma-ray flux during a rotation versus proximity to a gamma-ray flare peak), the
definition and fitting procedure of a gamma-ray flare, use of the $\Delta \theta_{\rm max}$ and $|\tau_{\rm obs}|$ CDFs
as test statistics and so forth. Making these choices introduces unavoidable and unquantifiable biases in our final results.
However, our exploratory analysis of the first-year data presented here has allowed us to identify well-defined statistical
questions, which we can address in a robust, a priori fashion using our second- and third-year data.

\section*{Acknowledgements}

The RoboPol project is a collaboration between Caltech in the USA,
MPIfR in Germany, Toru\'{n} Centre for Astronomy in Poland, the University of
Crete/FORTH in Greece, and IUCAA in India.
The University of Crete group acknowledges support by the ``RoboPol'' project, which is implemented under
the ``Aristeia'' Action of the  ``Operational Programme Education and Lifelong Learning'' and is
co-funded by the European Social Fund (ESF) and Greek National Resources, and by the European
Commission Seventh Framework Programme (FP7) through grants PCIG10-GA-2011-304001 ``JetPop'' and
PIRSES-GA-2012-31578 ``EuroCal''.
This research was supported in part by NASA grant NNX11A043G and NSF grant AST-1109911, and by the
Polish National Science Centre, grant number 2011/01/B/ST9/04618.
K.-\,T.- acknowledges support by the FP7 through
the Marie Curie Career Integration Grant PCIG-GA-2011-293531 ``SFOnset''.
M.-\,B.- acknowledges support from NASA Headquarters under the NASA Earth and Space Science Fellowship Program, grant NNX14AQ07H.
T.-\,H.- was supported in part by the Academy of Finland project number 267324.
I.-\,M.- and S.-\,K.- are supported for this research through a stipend from the International Max Planck
Research School (IMPRS) for Astronomy and Astrophysics at the Universities of Bonn and Cologne. 
The \textit{Fermi} LAT Collaboration acknowledges generous ongoing support from a number of agencies and institutes that
have supported both the development and the operation of the LAT as well as scientific data analysis. These include the
National Aeronautics and Space Administration and the Department of Energy in the United States, the Commissariat \`a
l'Energie Atomique and the Centre National de la Recherche Scientifique / Institut National de Physique Nucl\'eaire et de
Physique des Particules in France, the Agenzia Spaziale Italiana and the Istituto Nazionale di Fisica Nucleare in Italy,
the Ministry of Education, Culture, Sports, Science and Technology (MEXT), High Energy Accelerator Research Organization
(KEK) and Japan Aerospace Exploration Agency (JAXA) in Japan, and the K.~A.~Wallenberg Foundation, the Swedish Research
Council and the Swedish National Space Board in Sweden, Istituto Nazionale di Astrofisica in Italy and the Centre National
d'\'Etudes Spatiales in France.

%%%%%%%%%%%%%%%%%%%%%%%%%%%%%%%%%%%%%%%%%%%%%%%%%%



%\bibliographystyle{mnras}

\bibliographystyle{mnras}
% Use the LaTeX power, use bibtex properly.
\bibliography{bibliography_manual}

%%%%%%%%%%%%%%%%%%%%%%%%%%%%%%%%%%%%%%%%%%%%%%%%%%

%%%%%%%%%%%%%%%%%%%%%%%%%%%%%%%%%%%%%%%%%%%%%%%%%%

% Don't change these lines
\bsp	% typesetting comment
\label{lastpage}
\end{document}